\documentclass[aps, twocolumn, nobibnotes, superscriptaddress, notitlepage, noeprint,prapplied]{revtex4-2}
\usepackage[usenames,dvipsnames]{xcolor}
\usepackage{hyperref}
\hypersetup{linktoc=page,
colorlinks,
urlcolor=RedViolet,
citecolor=RedViolet,
linkcolor=RedViolet}

\usepackage{float}
\usepackage{graphicx}
\usepackage{mathrsfs}
\usepackage{overpic}
\usepackage{wrapfig}
\usepackage{braket}
\usepackage{color}
\usepackage{verbatim}
\usepackage{amssymb,amsmath,mathtools}
\usepackage{upgreek}
\usepackage{bbold}
\usepackage{cancel}
\usepackage{textgreek}
\usepackage[normalem]{ulem}
\usepackage{appendix}

\newcommand{\specialcell}[2][c]{%
  \begin{tabular}[#1]{@{}c@{}}#2\end{tabular}}

\begin{document}
\preprint{APS/123-QED}

\title{Machine Learning for Improved Current Density Reconstruction from 2D Vector Magnetic Images}

\author{Niko R. Reed}
\altaffiliation{Indicates co-first authors}
\affiliation{Department of Physics, University of Maryland, College Park, MD, USA}
\affiliation{Quantum Technology Center, University of Maryland, College Park, MD, USA}
\affiliation{Joint Quantum Institute, University of Maryland, College Park, MD, USA}

\author{Danyal Bhutto}
\altaffiliation{Indicates co-first authors}
\affiliation{Department of Biomedical Engineering, Boston University, Boston, MA, USA}
\affiliation{A. A. Martinos Center for Biomedical Imaging,
Massachusetts General Hospital, Charlestown, MA, USA}

\author{Matthew J. Turner}
\affiliation{Department of Physics, University of Maryland, College Park, MD, USA}
\affiliation{Quantum Technology Center, University of Maryland, College Park, MD, USA}
\affiliation{Department of Electrical and Computer Engineering, University of Maryland, College Park, MD, USA}

\author{Declan M. Daly}
\affiliation{Department of Physics, University of Maryland, College Park, MD, USA}
\affiliation{Quantum Technology Center, University of Maryland, College Park, MD, USA}

\author{Sean M. Oliver}
\affiliation{The MITRE Corporation, McLean, VA 22102, United States of America}

\author{Jiashen Tang}
\affiliation{Department of Physics, University of Maryland, College Park, MD, USA}
\affiliation{Quantum Technology Center, University of Maryland, College Park, MD, USA}

\author{Kevin S. Olsson}
\affiliation{Quantum Technology Center, University of Maryland, College Park, MD, USA}
\affiliation{Department of Electrical and Computer Engineering, University of Maryland, College Park, MD, USA}
\affiliation{Intelligence Community Postdoctoral Research Fellowship Program, University of Maryland, College Park, Maryland 20742, USA}

\author{Nicholas Langellier}
\altaffiliation{Present address: VideaHealth Inc., Boston, MA}
\affiliation{Department of Physics, University of Maryland, College Park, MD, USA}
\affiliation{Department of Physics, Harvard University, Cambridge, MA, USA}

\author{Mark J.H. Ku}
\altaffiliation{Present address: Northrop Grumman Mission Systems, Linthicum, Maryland 21090, USA.}
\affiliation{Department of Materials Science and Engineering, University of Delaware, Newark, DE 19716,
USA}
\affiliation{Department of Physics and Astronomy, University of Delaware, Newark, DE 19716, USA}

\author{Matthew S. Rosen}
\affiliation{A. A. Martinos Center for Biomedical Imaging,
Massachusetts General Hospital, Charlestown, MA, USA}
\affiliation{Department of Physics, Harvard University, Cambridge, MA, USA}
\affiliation{Harvard Medical School, 25 Shattuck St., Boston, MA, USA}

\author{Ronald L. Walsworth}
\email{walsworth@umd.edu}
\affiliation{Department of Physics, University of Maryland, College Park, MD, USA}
\affiliation{Quantum Technology Center, University of Maryland, College Park, MD, USA}
\affiliation{Department of Electrical and Computer Engineering, University of Maryland, College Park, MD, USA}

\date{December 12th, 2024}

\begin{abstract}
The reconstruction of electrical current densities from magnetic field measurements is an important technique with applications in materials science, circuit design, quality control, plasma physics, and biology. Analytic reconstruction methods exist for planar currents, but break down in the presence of high spatial frequency noise or large standoff distance, restricting the types of systems that can be studied. Here, we demonstrate the use of a deep convolutional neural network for current density reconstruction from two-dimensional (2D) images of vector magnetic fields acquired by a quantum diamond microscope (QDM). Trained network performance significantly exceeds analytic reconstruction for data with high noise or large standoff distances. This machine learning technique can perform quality inversions on lower SNR data,  significantly reducing the data collection time and permitting reconstructions of weaker and three-dimensional current sources. 
\end{abstract}

\maketitle

\section{Introduction}

The study of electric current distributions through imaging of Oersted magnetic fields has supported applications in many areas of physics, engineering, and medicine, in part due to its non-invasive nature \cite{orozco_magnetic_2009}.  Such applications include assessment of integrated circuits \cite{orozco_magnetic_2009, nowodzinski_nitrogen-vacancy_2015,infante_new_2009,levine_backside_2020,kehayias_measurement_2022,turner_quantum_2020,turner_magnetic_2020}, microelectronics \cite{infante_magnetic_2009,oliver_vector_2021,felt_construction_2007}, batteries \cite{hu_battery_2019,hu_sensitive_2020,bason_non-invasive_2022,brauchle_direct_2021}, solar panels \cite{scholten_imaging_2022,kaufmann_evaluation_2021}, superconducting tape and wires \cite{chikumoto_characterization_2021,sakai_effect_2023}, mechanical joints \cite{bevington_non-destructive_2018}, fault analysis \cite{wikswo_magnetic_1996}, superconducting materials and qubits \cite{wu_characterization_2021,kirtley_fundamental_2010,marchiori_magnetic_2022}, exotic materials \cite{ku_imaging_2020, feng_helical_2022,palm_imaging_2022,bjorlig_current_2022}, new semiconductor technology \cite{basso_electric_2023,lillie_imaging_2019}, and eddy current imaging \cite{wickenbrock_eddy_2016}. In medicine, imaging of magnetic fields generated by currents in biological tissues allows detailed and unique study of the heart, brain, and skeletal muscles \cite{roth_biomagnetism_2023,leder_noninvasive_1998,merwa_solution_2005,hamalainen_magnetoencephalography---theory_1993, alvarez_biomagnetic_1990, knappe_microfabricated_2016,limes_portable_2020}.

To enable these studies, current density distributions confined to thin planes are traditionally calculated utilizing Fourier space relationships between the source current density and the measured magnetic field.  Despite its widespread usage, this ``Fourier Method" for the magnetic inverse problem (i.e. the current density reconstruction problem) cannot effectively address high-noise measurements \cite{roth_using_1989}; it also performs poorly when the standoff distance between the measured magnetic field and the source current density is greater than the feature size of the current distribution. Due to these limitations, the Fourier Method sometimes results in poor analytic reconstructions \cite{turner_magnetic_2020, roth_using_1989}.  These drawbacks most heavily impact its utility for systems with small features or restricted standoff distance, such as nondestructive quality control and failure analysis of intact semiconductor chips, imaging of wires through walls or device boundaries, and noninvasive measurements of biological systems, which commonly exhibit weak magnetic field signals and small, irregular features \cite{roth_biomagnetism_2023}.

Due to the poor reconstruction performance of the Fourier Method in the low 
signal-to-noise ratio (SNR) regime and/or at large standoff distance, many repeated magnetic field measurements are typically required to achieve sufficient SNR, significantly slowing the measurement process and leading to systematic errors related to drift in the measured system or instrumentation \cite{levine_principles_2019}. An alternative current reconstruction technique requiring fewer measurements could enable higher fidelity results, the study of dynamic processes in samples that would otherwise be immeasurable, and greatly increase the speed of device quality control applications.

\begin{figure*}
    \centering
    \includegraphics[width=1.0\textwidth]{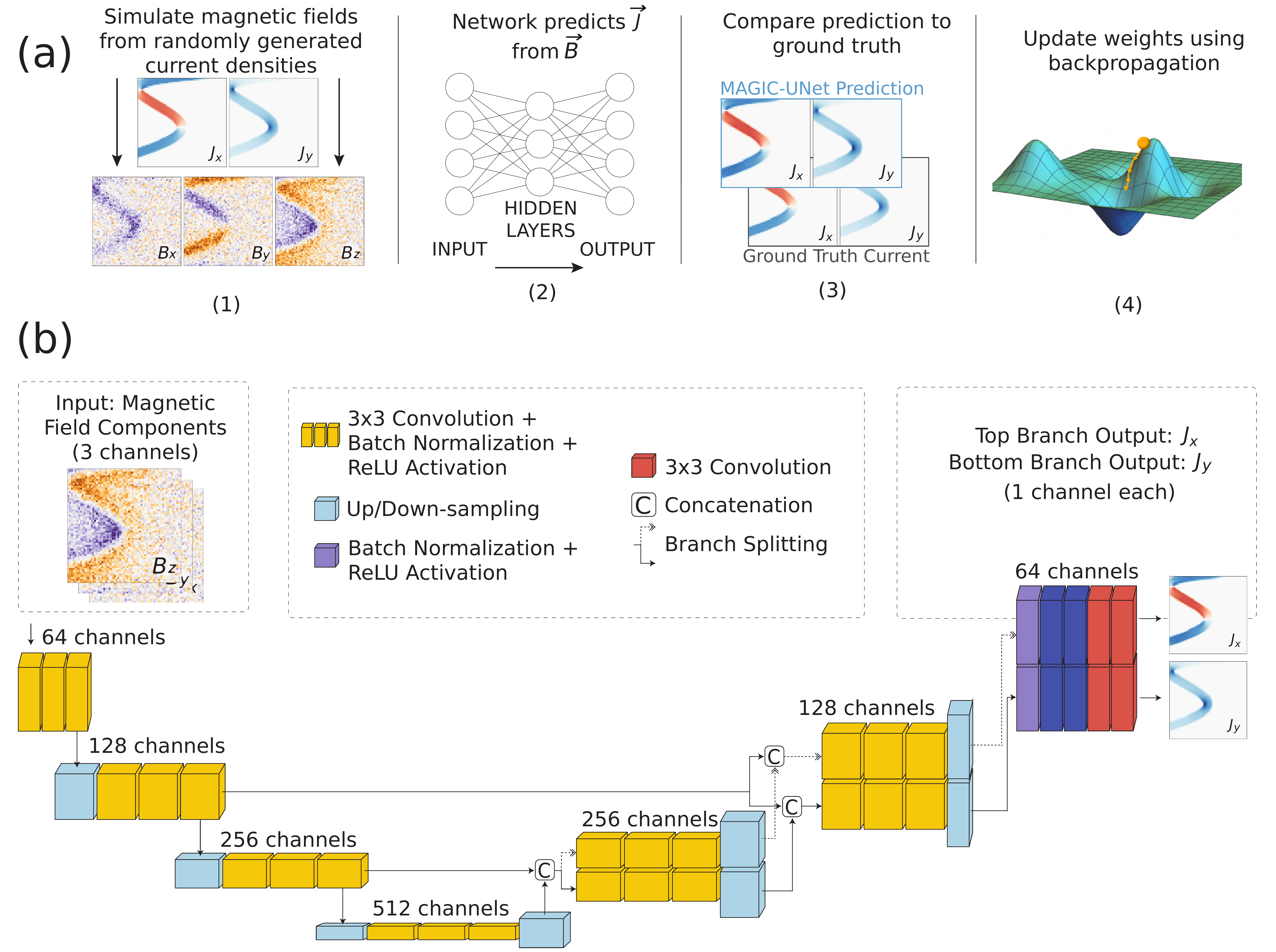}
    \caption{(a) Illustration of the training process. The MAGIC-UNet network is trained on field-current pairs, taking inputs of vector magnetic field images and making predictions of planar current density distributions. After each batch, the predicted currents are compared to the ground truth using mean squared error; and then network weights are updated to improve the accuracy of the prediction. After many iterations the network will have learned to make quality predictions for unknown currents. (b) Diagram showing the structure of the MAGIC-UNet network. Each colored block represents a layer of the given number of channels, which is indicated above the blocks. Downsampling uses a 2x2 average pooling layer; upsampling also includes a 3x3 convolutional layer. Concatenation is represented by a $\framebox{C}$  symbol and allows skip connections to join inputs from earlier parts of the network to later stages. The network has a branched structure with separate paths for $J_x$ and $J_y$, which do not interact.}
    \label{schematic}
\end{figure*}
\noindent

Neural networks excel at tackling a variety of image-based inverse problems \cite{jin_deep_2017,wang2020deep,schlemper2019dautomap,zhu_image_2018}. While early attempts at applying neural networks to the magnetic inverse problem operated in highly restricted parameter spaces \cite{coccorese_neural_1994,kishimoto_solution_1996}, advances in computing power and machine learning techniques show promise for identification of current density features of unrestricted sizes, shapes, and strengths in noisy images \cite{mccann_convolutional_2017}. 

Recently, a physically-informed network, trained to find the current density distribution that best reproduces the input magnetic field, has demonstrated success in low noise regimes, and can even can outperform the Fourier Method \cite{dubois_untrained_2022,tschudin_imaging_2024}. While this approach has the benefit of not requiring independent training data, it is not suitable for high noise inputs, as the network will seek to reproduce all input features including noise. A network trained with data pairs of noisy magnetic field inputs and associated current density distribution targets could allow the network to learn to differentiate signal from noise in the magnetic input signal. In particular, convolutional neural networks, including UNets \cite{ronneberger_u-net_2015} have shown impressive results in medical imaging reconstruction problems with regard to de-noising and artifact reduction  \cite{mehta2022mri,chen2017low}; and more generally in preserving structural features \cite{jin_deep_2017}. In addition to the demonstrated performance of UNet architectures, using a fully convolutional network reduces the computational demand of training with large images, and produces a network which can accept images of multiple resolutions. Both advantages stem from the fact that convolutional layers act by convoluting a filter kernal with the input data. Unlike linear layers, the size of the filter kernal is independent of the size of the input, allowing flexibilty in inference and enabling training with higher resolution data.

In this work, we present an adapted UNet architecture (MAGnetic Inverse Calculation UNet) designed for reconstructing current density images from vector magnetic field images with greater accuracy than the Fourier Method. The MAGIC-UNet network is tested on a variety of synthetic data types and experimentally collected Oersted magnetic field images using a Quantum Diamond Microscope (QDM). The QDM utilizes a dense, near-surface layer of nitrogen vacancy (NV) quantum defects in a diamond substrate to enable sensitive magnetic imaging, with micron-scale spatial resolution, via optical emission of the NV defects \cite{levine_principles_2019}. The MAGIC-UNet network is trained, in less than twelve hours, on hundreds of thousands of field-current image pairs generated analytically through Biot-Savart forward calculations of vector magnetic fields from randomly generated planar current density distributions (Figure \ref{schematic}a). We present current density reconstructions from both simulations and QDM experiments utilizing the MAGIC-UNet network, demonstrating greatly enhanced performance compared to the Fourier Method, and reliable reconstructions even in the presence of extreme noise (input SNR $\sim$ 1).

\section{Network Architecture}
Characteristic to UNet architectures, MAGIC-UNet combines low and high resolution image filtering through a series of convolutional layers \cite{ronneberger_u-net_2015}. The purpose of the network is to take a vector magnetic field image in a plane above the sample (expressed in components $B_x \text{, } B_y \text{ and } B_z$ for each pixel in the image), and output the associated planar current density distribution (components $J_x \text{ and } J_y$) for each point in the source plane. The network architecture is divided into a downsampling portion followed by a branched upsampling portion, with one branch for each component of current density. 

The MAGIC-UNet network begins image processing with a series of downsampling steps that reduce the image size and allow the network to capture low-level features \cite{jin_deep_2017}. During this process, the resolution of a given input image is reduced by a factor of 2 for each step while the number of channels is increased by 2. After 3 downscaling iterations, a similar upsampling process occurs that increases the resolution, ultimately producing final outputs of the same resolution as the starting images.  In addition, layers from the downsampling path are directly connected through skip connections to upsampled layers of the same resolution to bypass intermediate layers. These connections enable the preservation of structure in the data across multiple resolutions that would otherwise be lost due to downsampling. Skip connections also ameliorate the vanishing gradient problem by allowing gradients to flow to earlier layers before the gradient diminishes \cite{bengio1994learning}. We omit the first skip connection as we find it does not impact MAGIC-UNet performance.

During each iteration of the downsampling pathway, a padded 3x3 covolutional layer is applied, followed by feeding layers through a batch normalization layer and a Rectified Linear Unit (ReLU) activation function. This process is repeated three times before downsampling with a 2x2 average pooling layer with a stride of 1. Average pooling layers are used instead of max pooling layers, as all image details and features are important to our objective. When all downsampling is completed, the network forms two branches that upscale separately, processing $J_x \text{ and } J_y$, respectively. The network first performs a 2x2 upsampling convolution, then uses a skip connection to concatenate the channels from the downsampling pathway to the channels from the upsampling pathway of the same resolution. The process of applying a 3x3 covolutional layer, batch normalization, and ReLu activation occurs three times. Specific inter-layer connections and data resolutions at each step are depicted in Figure \ref{schematic}b.

\section{Results}

\begin{figure*}[ht]
    \centering
    \includegraphics[width=1.0\textwidth]{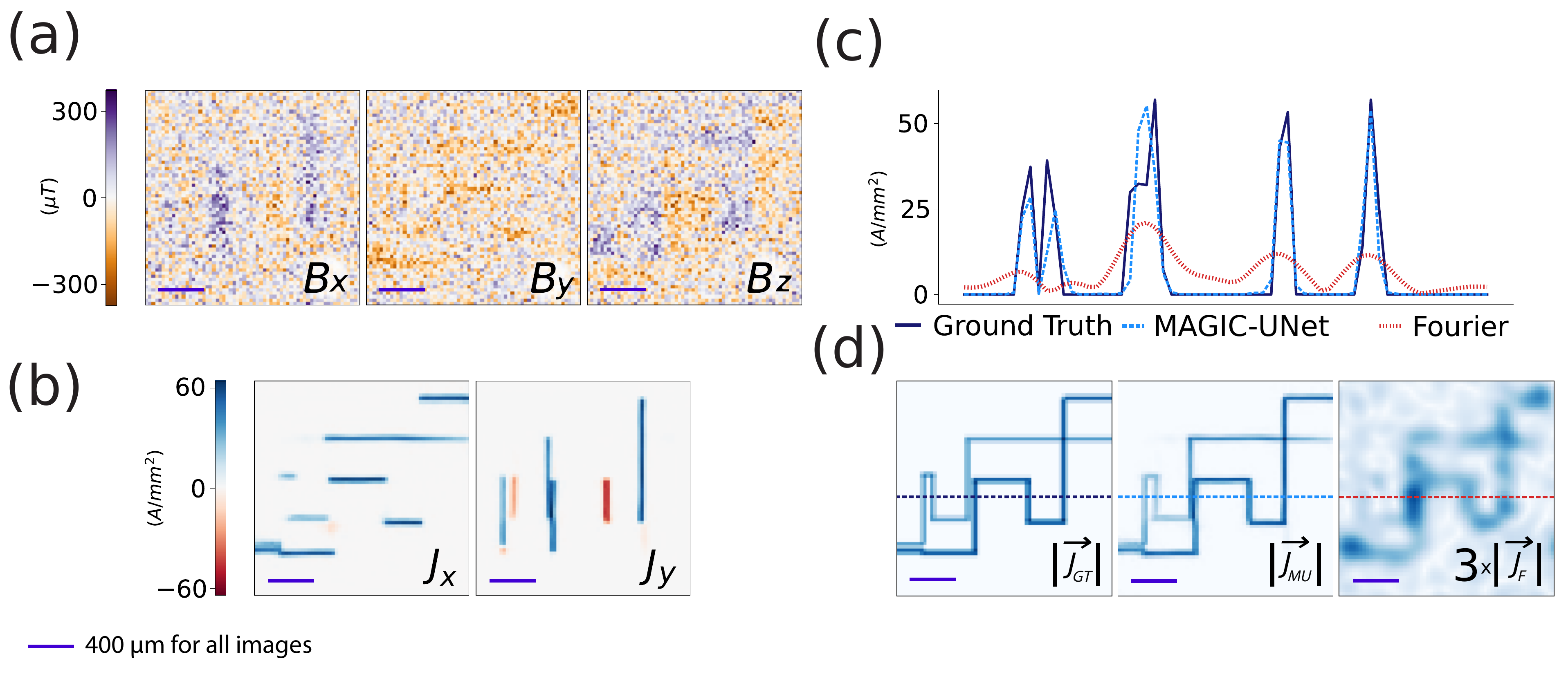}
    \caption{Current density predictions for synthetic (simulated) vector magnetic field images. The validation (input) data set includes magnetic field images in a plane separated by 50 ± 10 µm from the planar current density distribution. (a) x, y, and z components of an example input magnetic field image, including added Gaussian noise at $\sigma$=0.5. (b) MAGIC-UNet current density predictions for input shown in (a), with x and y components shown separately. (c) Linecuts showing the current density magnitude comparing the ground truth (black), MAGIC-UNet (blue), and Fourier Method (red). The location of the linecuts is indicated by the dashed lines in (d). (d) Current density magnitude showing (left to right) the ground truth, MAGIC-UNet, and Fourier Method predictions for input shown in (a). The Fourier Method image has been globally multiplied by 3 to allow it to be visible on the same scale as the other two images. Scale bars are 400 µm for all planar images shown in (a), (b), and (d).}
    \label{syn_data}
\end{figure*}

Using diverse validation datasets, the MAGIC-UNet network and Fourier Method are tasked with producing $J_{x, y}$ field inversions from vector magnetic field images both synthetic (Figure \ref{syn_data}a) and experimental; and the performance of the two methods is compared. See Tables \ref{table1} and \ref{table2}, as well as discussion in the Supplemental Information. The accuracy of the inversions is evaluated by comparing to ground truth current densities using both qualitative and quantitative methods. As discussed below, MAGIC-UNet outperforms the Fourier Method by all metrics across all types of magnetic field input data.

\begin{table}
\begin{ruledtabular}
\begin{tabular}{c|cccc}
& \specialcell{MAGIC- \\UNet $J_x$} & \specialcell{MAGIC- \\UNet $J_y$} & \specialcell{Fourier \\ Method $J_x$} & \specialcell{Fourier \\ Method $J_y$} \\
\hline
\specialcell{Right \\ Angle} & 0.895 & 0.898 & 0.421 & 0.421 \\
\hline
\specialcell{All \\ Angles} & 0.901 & 0.904 & 0.455 & 0.455 \\
\hline
\specialcell{Thick\\ Straight} & 0.900 & 0.902 & 0.359 & 0.359 \\
\hline
\specialcell{Thin \\Curves} & 0.962 & 0.962 & 0.356 & 0.358 \\
\hline
\specialcell{Thick \\Curves} & 0.950 & 0.947 & 0.396 & 0.399 \\
\hline
\specialcell{Out-of-\\Distribution} & 0.948 & 0.947 & 0.519 & 0.514 \\
\end{tabular}
\end{ruledtabular}
\caption{\label{table1}Average Structural Similarity Index Measure (SSIM) for the $J_x$ and $J_y$ channels of MAGIC-UNet and the Fourier Method for the five classes of data in the validation set, as well as a set of out-of-distribution data. The Right Angle, All Angles, and Thick Straight classes can contain 1, 2, or 3 closely spaced current planes, while the Thin Curves and Thick Curves form shorts where currents overlap. Uncertainties for all average SSIM values are very small ($\approx$ 0.0001).}
\end{table}

\begin{table}
\begin{ruledtabular}
\begin{tabular}{c|cccc}
& \specialcell{MAGIC- \\UNet $J_x$} & \specialcell{MAGIC- \\UNet $J_y$} & \specialcell{Fourier \\ Method $J_x$} & \specialcell{Fourier \\ Method $J_y$} \\
\hline
\specialcell{One \\ Plane} & 0.966 & 0.969 & 0.411 & 0.412 \\
\hline
\specialcell{Two \\ Planes} & 0.902 & 0.907 & 0.415 & 0.414 \\
\hline
\specialcell{Three\\ Planes} & 0.842 & 0.845 & 0.406 & 0.403 \\
\end{tabular}
\end{ruledtabular}
\caption{\label{table2} Average SSIM for the $J_x$ and $J_y$ channels of MAGIC-UNet and the Fourier Method, for the three data classes that contain multiple closely spaced current planes. Uncertainties for all average SSIM values are very small ($\approx$0.0001).}
\end{table}

We begin by assessing the performance of MAGIC-UNet and the Fourier Method on a synthetic validation data set, with standoff distances for each synthetic image sampled from a random distribution with mean 50$\,\mu$m and standard deviation 10$\,\mu$m to mimic the variation and uncertainty of an experimental dataset. While the exact standoff distance is used in the Fourier Method calculation, MAGIC-UNet is able to reconstruct accurate current densities without access to this information; a practical feature for experimental applications. 

Based on the visual fidelity of current density maps,  MAGIC-UNet predictions closely resemble ground truth $J_x$ and $J_y$.  The network performs notably well in addressing complex $\vec{J}$ features, including overlapping wire structures and sections with nonuniform current density (e.g., see Figure \ref{syn_data}b). Comparatively, the Fourier Method consistently produces diffuse noise artifacts, blurring, and substantial smoothing of current edges. In addition to visual features, the peak current density values predicted by MAGIC-UNet are more accurate than those of the Fourier Method, which predicts peak current densities 2-3 times smaller than the true values; an example of this behavior is shown in Figure \ref{syn_data}c. 

\noindent

For quantitative evaluations of inversion accuracy, we primarily employ the Structural Similarity Index Measure (SSIM), a metric used in image processing applications to examine preservation of structural information. SSIM compares the inversion to its ground truth. SSIM values of 0.95-0.99 are generally considered to have minor to imperceptible deviance from the true current density; values less than 0.88 are ``poor" quality; and values below 0.50 are ``bad" quality \cite{zanforlin_ssim-based_2014,flynn_image_2013}. Additionally, we expand our quantitative evaluation of inversion accuracy using Root Mean Square Error (RMSE) and Peak Signal to Noise Ratio (PSNR), which emphasize different traits of images.

To study the behavior of the MAGIC-UNet network and Fourier Method under variable levels of noise, we introduce artificial noise to the simulated input $\vec{B}$ data by applying an additive filter that adjusts pixel values by random amounts sampled from a Gaussian distribution.  In our analysis, we define the noise level $\sigma_n$ of a magnetic field image as the ratio of the magnetic field noise standard deviation $\sigma$ to the maximum value of the magnetic field magnitude $\text{max}(\big|\vec{B}\big|)$, across all image pixels. This normalization controls the Normal Distribution $\mathcal{N}(\mu=0, \sigma_n^2)$ of the additive filter applied to the image.  For example, a noise level of $\sigma_n = 0.2$ corresponds to setting the standard deviation $\sigma$ of the distribution $\mathcal{N}$ used in the additive filter process to one-fifth of the maximum magnetic field magnitude in the image. In general, we find accurate predictions from MAGIC-UNet for $\sigma_n=0.5$, roughly corresponding to an input image data SNR of 2. MAGIC-UNet is typically able to identify current structures for noise levels as large as $\sigma_n=1.0$ (input SNR $\sim$ 1). However, network predictions are increasingly distorted as the noise level is increased; examples are shown in Figure \ref{noise}a,b.

\subsection{Quantitative Performance}

\begin{figure*}
    \centering
    \includegraphics[width=1.0\textwidth]{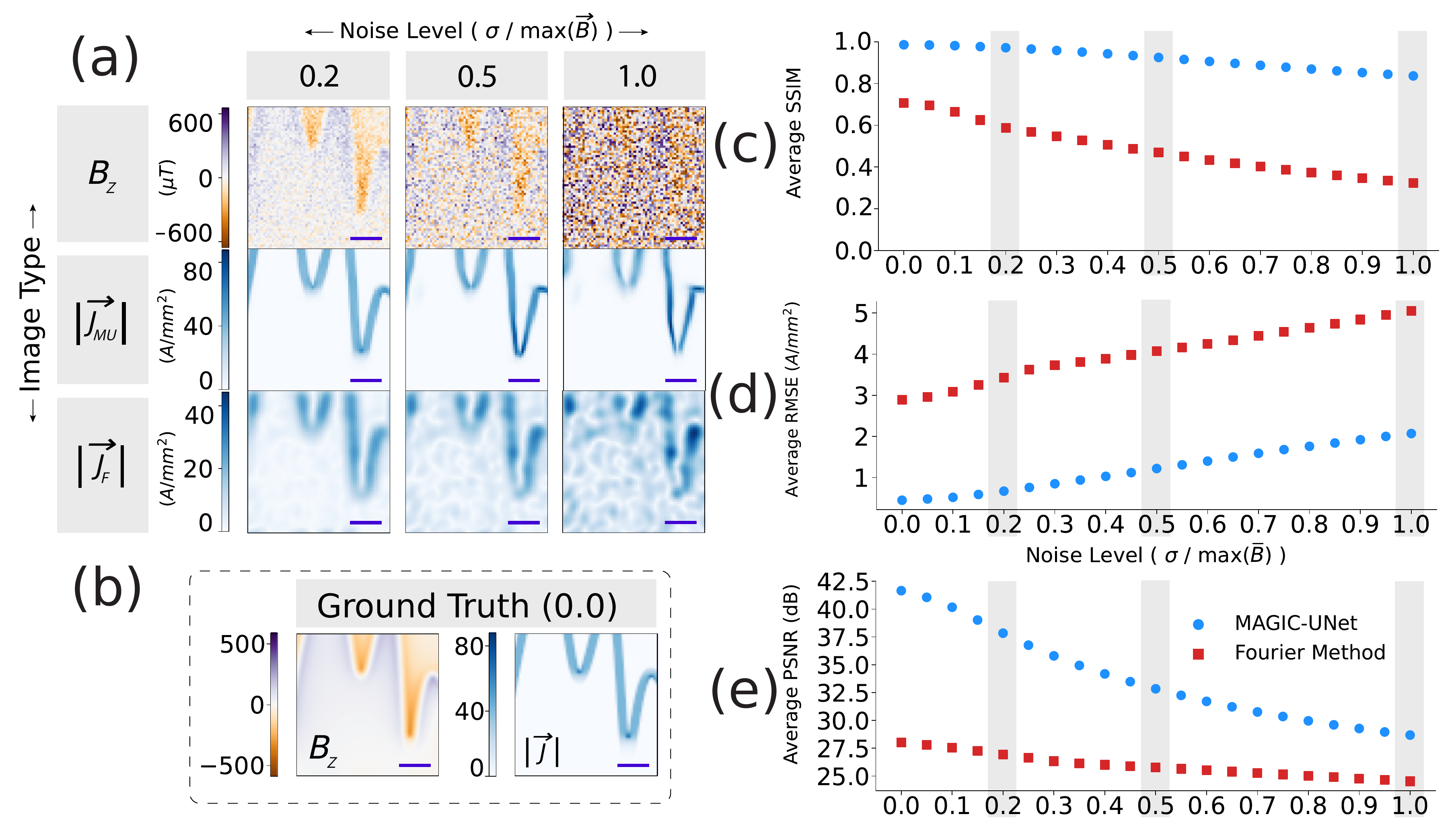}
    \caption{Comparison of MAGIC-UNet and Fourier Method current density reconstructions for simulated input magnetic imaging data, with varying noise.  The validation (input) data set includes magnetic field images in a plane separated by 40 ± 5 µm from the planar current density distribution. (a) z-component of an example input magnetic field image with added noise; and resulting magnitude of current density predictions for MAGIC-UNet and the Fourier Method. Three levels of noise are shown, for the same underlying ground truth, corresponding to $\sigma_n=$ 0.2, 0.5, and 1.0. (b) Ground truth z-component of magnetic field image and current density magnitude. (c) Average SSIM for MAGIC-UNet (blue circles) and the Fourier Method (red squares) determined from validation data with different values of noise. (d) Average RMSE across the validation data for MAGIC-UNet and the Fourier Method. (e) Average PSNR across the validation data for MAGIC-UNet and the Fourier Method. Gray boxes in (c)-(e) correspond to noise levels displayed in (a). Scale bars are 400 µm for all planar images shown in (a) and (b).}
    \label{noise}
\end{figure*}

The quantitative performance of MAGIC-UNet supports its visual fidelity.  When inverting validation $\vec{B}$ data and comparing to the ground truth current density $\vec{J}_{GT}$, MAGIC-UNet achieves SSIM values above 0.95 (corresponding to good image quality) when given validation data with noise levels $\sigma_n$ between 0 and 0.35  (input SNR $>$ 2.8)  ; and SSIM values above 0.88 (corresponding to "fair" image quality) for $\sigma_n$ between 0.40 and 0.70 (2.8 $>$ input SNR $>$ 1.4); see Figure \ref{noise}c). Overall, SSIM values range from 0.99 at $\sigma_n=0$ to 0.84 at $\sigma_n=1$. In contrast, SSIM values for Fourier Method predictions span 0.71 to 0.32 over the same range, classifying as ``poor" between noise levels of $\sigma_n=0.0$ to $\sigma_n=0.4$  (input SNR $>$ 2.5) , and as ``bad" for $\sigma_n > 0.45$.  Throughout these tests, the MAGIC-UNet network does not favor any particular current density component: the difference in average SSIM for $J_x$ and $J_y$ does not exceed 1\% of their total values, performing similarly in this regard to the Fourier inversions (see Tables \ref{table1} and \ref{table2}).  

SSIM performance is similar across all types of in-distribution data, with the exception that distributions with multiple current planes lead to significantly worse average SSIM values for MAGIC-UNet predictions (Table \ref{table2}). To check for overfitting, we compare these results to the SSIM of reconstructions of out-of-distribution validation data. This out-of-distribution data, primarily consisting of straight and curved wire segments, is generated using a separate algorithm than for training and validation data, with corresponding magnetic fields simulated using COMSOL Multiphysics\textsuperscript{\textregistered} instead of the analytic process described in Section \ref{Methods}. (Also see discussion in the Supplemental Information.) For the example out-of-distribution data, MAGIC-UNet achieves SSIM of 0.962 $\sigma_n=0.5$, which is within the range of the in-distribution single-layer datasets.  However, when tasked with performing inversions on test data far outside of the distribution, such as intricate, branching wires with non-uniform width, MAGIC-UNet achieves a SSIM value of 0.88 at noise level $\sigma_n=0.5$, showing that the network has not learned a fully general version of the inversion and may be biased towards distributions more similar to the training data sets. Interestingly, the gap between performance on in- and out-of-distribution validation data disappears at the highest noise levels, suggesting these effects are more important for ensuring high fidelity predictions given mild to moderate noise than for approximating the general features in a high noise image.

MAGIC-UNet predictions also exhibit less average error as indicated by the RMSE and PSNR metrics. Mean Square Error (MSE) and the related RMSE are frequently used to evaluate regression predictions, as MSE is the maximum likelihood estimator for data with Gaussian noise, such as our experimental data, and served as the loss function during training \cite{hodson_root-mean-square_2022}. The average RMSE of Fourier Method predictions is about 3-4x greater than the MAGIC-UNet RMSE at every level of noise tested (Figure \ref{noise}d). PSNR is perhaps the most widely used metric in image and video processing and seeks to quantify the ratio between reconstruction erors and the true signal \cite{Winkler_Evolution_2008}. PSNR for MAGIC-UNet predictions ranges from 41.7 dB at $\sigma_n = 0$ to 28.7 dB at $\sigma_n = 1$. Comparatively, PSNR for Fourier Method predictions ranges from 28.0 dB to 24.5 dB over the same interval (Figure \ref{noise}e). The superior RMSE and PSNR for MAGIC-UNet relative to the Fourier Method is manifested by both fewer noise artifacts and more accurate current density predictions.

\noindent

\begin{figure}
    \centering
    \includegraphics[width=0.5\textwidth]{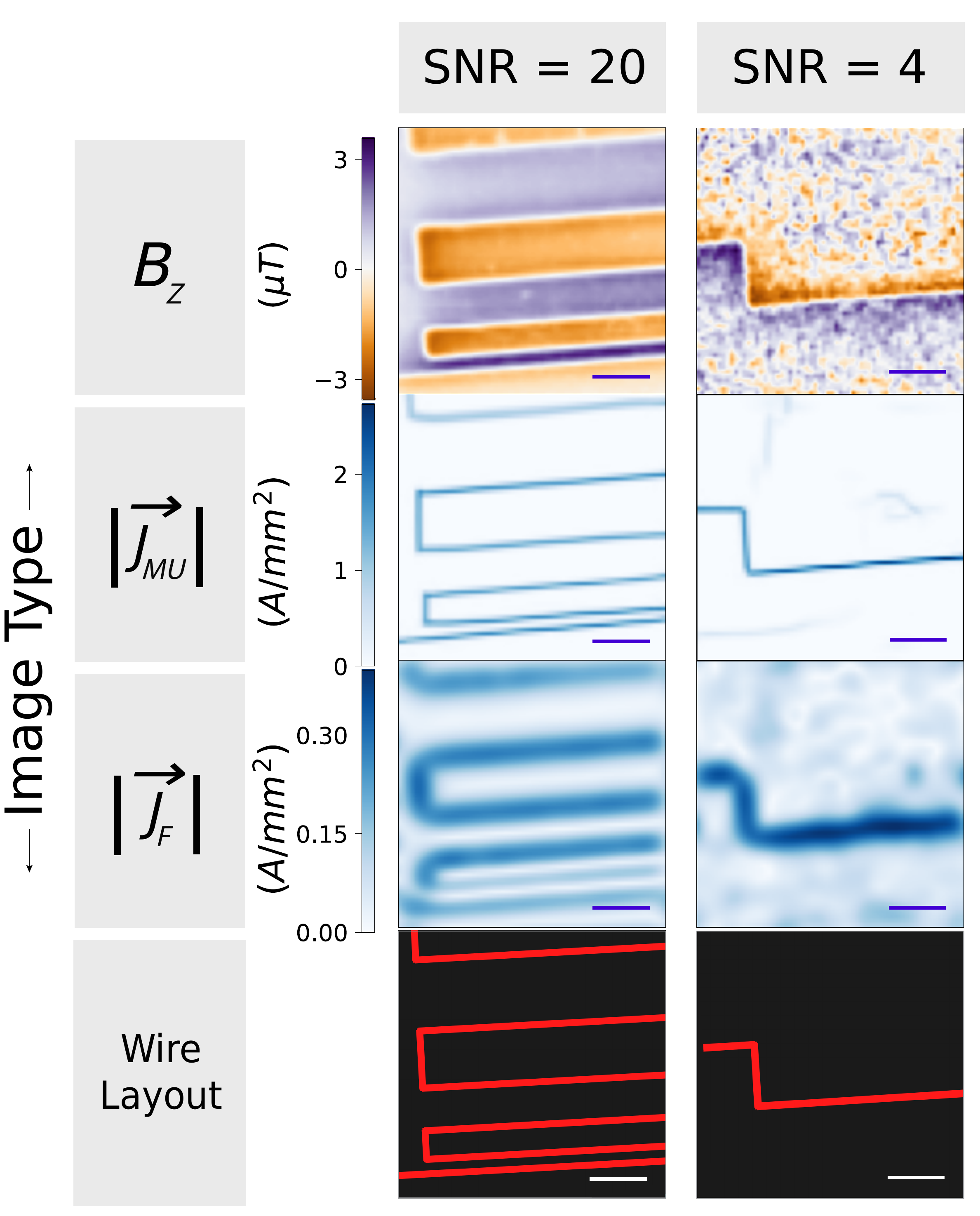}
    \caption{Current inversion for two levels of SNR on experimental (QDM) magnetic field imaging data. The magnetic field images are in a plane separated from the current source by 65$\,\mu$m (left column) and 36$\,\mu$m (right column). Rows show the z-component of measured magnetic field images; the current density magnitudes from MAGIC-UNet and Fourier Method reconstructions; and the wire layout used to generate the source current for magnetic field images. Low SNR data is collected from a single-shot QDM image; while high SNR data is  created from an average of 400 QDM images of the same magnetic field pattern. The wire layouts are cropped and rotated to match the QDM field of view. Scale bars are 400 µm for all planar images.}
    \label{experimental}
\end{figure}

When tasked with inverting experimental data collected using a QDM, MAGIC-UNet shows similar performance improvements over the Fourier Method as seen with synthetic data, including fewer noise artifacts and more accurate wire widths.  As an example, the predicted wire widths in Figure \ref{experimental} are visibly much more accurate in MAGIC-UNet reconstructions than in Fourier Method reconstructions, when compared with experimental QDM magnetic images and wire layout ground truth. 

However, more noise artifacts appear in MAGIC-UNet reconstructions of experimentally collected images than in reconstructions of synthetic images with the same $\sigma_n$. This result is likely due to spatially-correlated noise in the experimental images due to the fitting process for NV Optically Detected Magnetic Resonance (ODMR) spectra; whereas the MAGIC-UNet network is trained on simulated magnetic field images with uncorrelated Gaussian noise. See discussion in the Supplemental Information. 

\noindent

\subsection{Increasing Spatial Resolution}
\begin{figure*}
    \centering
    \includegraphics[width=1.0\textwidth]{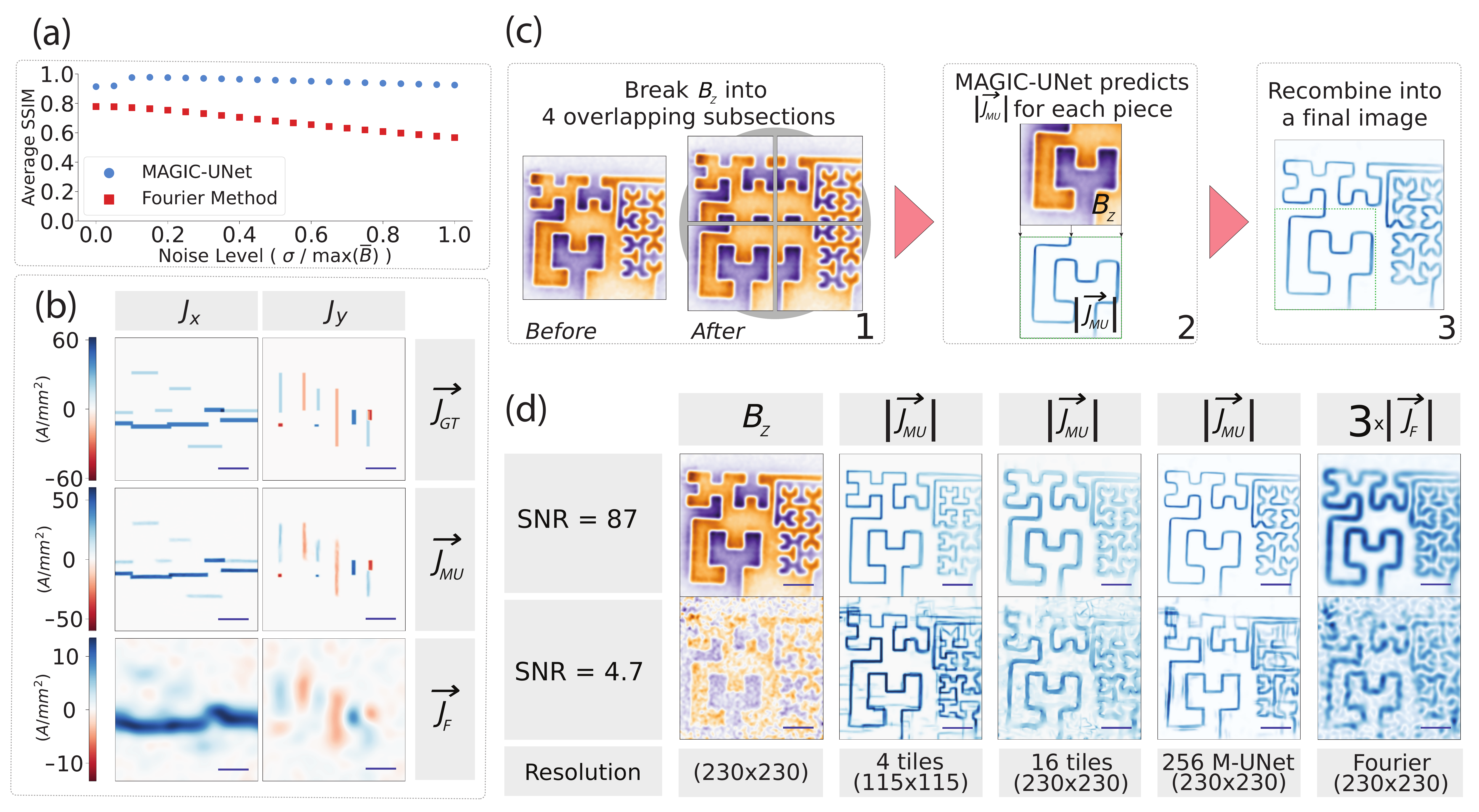}
    \caption{Assessment of methods for efficient current density reconstruction of higher spatial resolution images than the (nominal) 64x64 resolution MAGIC-UNet network.  (a) Comparison of average SSIM for 256$\times$256 MAGIC-UNet and Fourier Method for synthetic data of varying $\sigma_n$. The synthetic magnetic field images are in a plane separated by 50 ± 10 μm from the planar current density distributions. (b) Comparison of current density reconstructions for (top to bottom) ground truth, 256$\times$256 MAGIC-UNet, and Fourier Method on synthetic data with $\sigma_n =$ 0.5 added noise. Columns show $J_x$ and $J_y$. (c) Work flow of tiling process. (1) A magnetic field image is divided into overlapping 64x64 sections. (2) Each section is passed separately to the reconstruction protocol (MAGIC-UNet or Fourier Method). (3) The full reconstruction is stitched together from the individual predictions. Overlapping  regions are divided equally between tiles, with duplicate predictions discarded. (d) Comparison of current density predictions using 4 different methods for both high and low SNR experimental data. QDM $B_z$ images created by a Hilbert curve wire array are shown in column 1 with a resolution of 230x230 pixels. Current density reconstruction methods include a 4 tile implementation of 64x64 MAGIC-UNet producing 115x115 images (column 2); a 16 tile implementation of 64x64 MAGIC-UNet producing 230x230 images (column 3); a cropped version of (untiled) 256$\times$256 MAGIC-UNet (column 4); and a 230x230 resolution version of the Fourier Method (column 5). The QDM magnetic field images are in a plane separated by 45$\,\mu$m from the current source.  Scale bars are 400 µm for all planar images shown in (b) and (d).}
    \label{tiling}
\end{figure*}

To facilitate inversions featuring spatially detailed current profiles, we assess two methods of increasing the input spatial resolution of the network. While the MAGIC-UNet model does not explicitly restrict the input size, the performance of the network degrades as the given input diverges from the image resolution on which the network is trained.  Thus, we train a separate network using 256$\times$256 data and a similar model, but with an additional downsampling and upsampling layer due to the larger input size. The resulting 256$\times$256 resolution MAGIC-UNet performs well on inversions of synthetic data, outperforming its (nominal resolution) 64x64 counterpart for high $\sigma_n$ and the 256$\times$256 resolution Fourier Method at all $\sigma_n$ (see Table \ref{restable} and Figure \ref{tiling}a,b). The 256$\times$256 MAGIC-UNet network also performs well on low noise QDM experimental data, but degrades in performance compared to its 64x64 counterpart on higher-noise experimental data. This discrepancy may be explained by experimental image artifacts being significantly less Gaussian at 256$\times$256 resolution than in a 64x64 pixel image, which is created through binning of the raw QDM experimental image.

Although the 256$\times$256 MAGIC-UNet network performs well with the exception of high-noise experimental data, the enhanced resolution substantially increases the memory and computational time needed for training. To alleviate this shortcoming, we implement a tiling method that can effectively process larger images, with each image region employing the 64x64 resolution MAGIC-UNet network (Figure \ref{tiling}c). In this process, higher resolution images are segmented into overlapping 64x64 pixel tiles. MAGIC-UNet then processes each tile as if it is a separate data set.  Once complete, overlapping tiles are trimmed and recombined into an image with the same resolution as the starting input. We test both 4 and 16 tile configurations on experimental data, allowing for resolution upscaling of approximately 2 and 4 times the linear size respectively.

As shown in Figure \ref{tiling}d, MAGIC-UNet makes satisfactory predictions using this tiling method, with superior SSIM to the full resolution Fourier Method for all noise levels. Note that the performance of the (tiled) MAGIC-UNet network is for a field of view up to four times smaller than the network training data, which is an interesting result to be studied in future work, given that the field of view significantly impacts the analytic transformation underlying the network architecture. The success of the tile predictions demonstrates that the MAGIC-UNet network can handle data outside of the parameter range on which it was trained. Further work could investigate whether training a network specifically to handle smaller fields of views for a tiling application leads to even better performance. 

\begin{table}[H]
\begin{ruledtabular}
\begin{tabular}{c|cccc}
& \specialcell{SSIM\\ $\sigma_n=0.1$} & \specialcell{SSIM\\$\sigma_n=0.2$} & \specialcell{SSIM\\$\sigma_n=0.5$} & \specialcell{SSIM \\ $\sigma_n=1.0$} \\
\hline
\specialcell{64x64\\ MAGIC-UNet} & 0.981& 0.971& 0.925& 0.836\\
\hline
\specialcell{256$\times$256\\MAGIC-UNet}& 0.976& 0.975 & 0.958 & 0.923\\
\hline
\specialcell{256$\times$256\\ Fourier Method} & 0.771&  0.754 & 0.680 & 0.567\\
\end{tabular}
\end{ruledtabular}
\caption{\label{restable} Average SSIM for 64x64 and 256$\times$256 MAGIC-UNet and 256$\times$256 Fourier Method for inversions of synthetic data with different $\sigma_n$.}
\end{table}

\noindent

\subsection{Large Standoff}
 
\begin{figure*}
    \centering
    \includegraphics[width=1.0\textwidth]{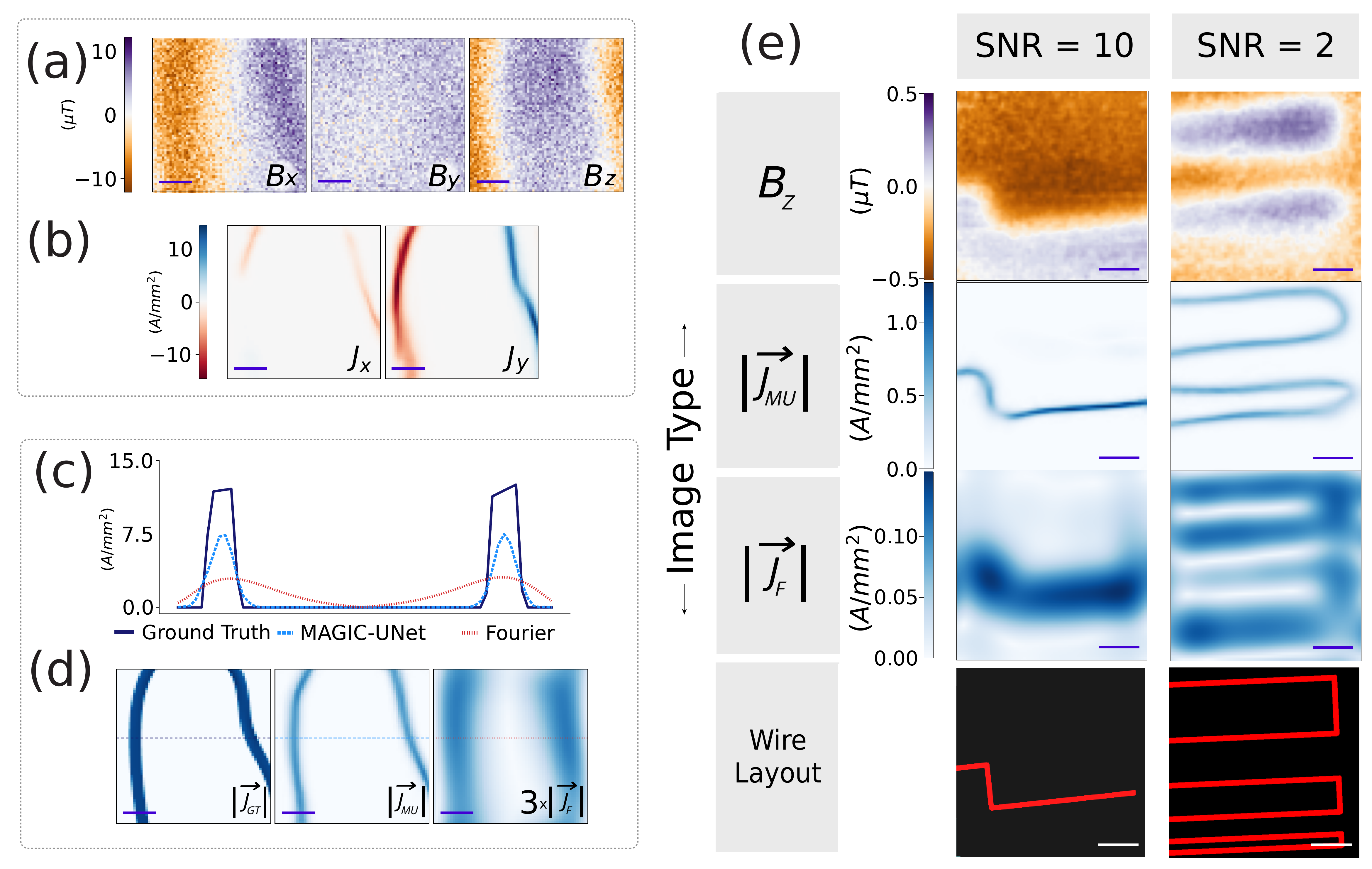}
    \caption{Assessment of current density reconstruction methods at large standoff distances. (a) Images of x, y, and z components of magnetic field for synthetic data at 500 $\mu$m standoff with $\sigma_n =$ 0.2. (c) $J_x$ and $J_y$ predictions from MAGIC-UNet for input data in (a). (c) Linecuts comparing current density for ground truth (black), MAGIC-UNet (blue), and Fourier Method (red). (d) Comparison of ground truth current density magnitude image with MAGIC-UNet and Fourier Method predictions. Fourier Method prediction is shown at 3x magnitude in order for it to be visible in the same data range as the other two plots. Dashed lines show the location of linecuts plotted in (c). (e) QDM $B_z$ images, MAGIC-UNet $\big|\vec{J}\big|$, and Fourier Method $\big|\vec{J}\big|$ for high-SNR and low-SNR experimental data taken at standoff distance $\approx$ 500 $\mu$m. Scale bars are 400 µm for all planar images shown.}
    \label{far}
\end{figure*}

To examine the large standoff regime, we test a 64x64 (untiled) MAGIC-UNet network trained on magnetic field data simulated at a standoff distance of 500 ± 50 $\mu$m above the current plane, in contrast to the standoff distance of 50 ± 10 $\mu$m used in all other synthetic data in this work. This 500 $\mu$m distance corresponds to about 5 to 10 times the current source feature size and is therefore well beyond the regime where the Fourier Method is expected to perform well. Example results are shown in Figure \ref{far}a-d.

In this challenging regime, both reconstruction methods perform too poorly for our quantitative metrics to provide meaningful insight. However, it can be observed qualitatively that the MAGIC-UNet network performs better than the Fourier Method.  For example, at $\sigma_n = 0.2$, Fourier Method predictions are severely hindered by input image blurring and edge effects, yielding spatially ill-defined current reconstructions in the general area of the current source. In comparison, MAGIC-UNet predicts spatially well-defined wires of roughly correct size, albeit with notably increased artifacts and inaccuracies compared to input data from a 50$\,\mu$m standoff distance. Predictions on experimental data show a similar pattern: MAGIC-UNet makes more meaningful predictions for source current density than the Fourier Method due to its ability to yield current sources of reasonable width flowing in similar directions to the ground truth.  Notable deviations of MAGIC-UNet from ground truth include rounded corners for wires, as visible in Figure \ref{far}e.

\noindent

\section{Discussion}

The above results illustrate the utility of MAGIC-UNet for high-fidelity reconstruction of planar current density distributions, whether for synthetic or experimental data, from two-dimensional images of vector Oersted magnetic fields.  Performance of the MAGIC-UNet network, both qualitatively and quantitatively, is consistently superior to that of a conventional Fourier reconstruction method, including for cases that degrade the fidelity of both approaches: e.g., reduced input data SNR  (a regime that also reduces the performance of physically informed networks \cite{dubois_untrained_2022}). The improvement is quite significant; e.g. applied to 50$\,\mu$m standoff data, the 64x64 network performs about 10$\times$ better than the Fourier Method by RMSE with no noise, and about 2.5$\times$ better at SNR = 1.  MAGIC-UNet performs better in SSIM, RMSE, and PSNR at $\sigma_n = 1$ than the Fourier Method does at $\sigma_n = 0$; showing the capacity to dramatically reduce measurement time which would otherwise be required to produce high SNR images for Fourier Method reconstruction. Using SNR=1 inputs for MAGIC-UNet reconstruction, which we found to be more accurate than input-SNR=20 Fourier Method reconstructions, would allow data to be collected 400 times faster.

With the ability to utilize faster experimental readout and associated weaker input signals, we are optimistic MAGIC-UNet could be a preferred alternative to the Fourier Method for current density reconstruction from magnetic imaging data.

A key challenge for future work will be assessing the utility of the MAGIC-UNet architecture for reconstruction of three-dimensional (3D) current sources from vector magnetic field images. Though the unconstrained (general) magnetic inverse problem does not have a unique solution, there are 3D solutions for specific cases of interest, e.g., when the current is constrained to discrete wires \cite{hauer_uniqueness_2005}. An intermediate step that may first be investigated is a circuit with current constrained to multiple planes plus occasional wires between planes. Current reconstruction of such a "2.5D" circuit might first separate magnetic field images associated with distinct planes before determining the associated current densities.  $\vec{J}$ data could be handled similarly to the approach of MAGIC-UNet through an iterative process removing the topmost layer, or differently as a single step that performs the entire transformation for the entire multiple-plane circuit. Finally, another challenge for future work is to adapt the MAGIC-UNet methodology to magnetization distributions or combined current and magnetization distributions.

\section{Methods}
\label{Methods}
\subsection{Synthetic Data Creation and Training}

We developed a data generation workflow to create 153,600 synthetic $\vec{J}$-$\vec{B}$ current-field pairs for model training and 4096 field-current pairs for validation and testing.  Starting with random shapes produced by multiple wire-generation algorithms, finite element simulations in both COMSOL Mutliphysics\textsuperscript{\textregistered} and MATLAB calculate the planar current density resulting from a voltage difference across the boundary of the simulated conductive material. To ensure robust training, we include a wide variety of wires, including shorts, closely spaced current layers, and non-uniform current density around corners.  Once the current density simulations are complete, Oersted magnetic fields are calculated analytically using the Biot-Savart law. Additional details regarding training data generation are provided in the Supplementary Information.

The RMSProp optimization algorithm \cite{hinton2018lecture} is employed for training the network. Training details are documented in Table \ref{training_parameters_table}. The approach to data creation and training does not require access to specialized equipment or extensive computational time, allowing models to be trained on demand for different needs. The entire process from creating the training data to training the model requires about 24 hours. Once the network is trained, current density predictions can be generated for thousands of magnetic field images in a few seconds, making the method extremely efficient.

\begin{table}[H]
\begin{ruledtabular}
\begin{tabular}{cccccc}
\specialcell{Learning \\ Rate} & Epochs & \specialcell{Mini-\\batch} & \specialcell{Loss \\ Function} & Library & \specialcell{Graphics \\ Processor}\\
\hline
0.0001 & 50 & 32 & MSE & \specialcell{TensorFlow \\ 2.10.0} & \specialcell{NVidia \\ RTX-A6000}\\
\end{tabular}
\end{ruledtabular}
\caption{\label{training_parameters_table}
MAGIC-UNet training details using RMSProp optimization algorithm. The learning rate is set to 0.0001, while the training process utilizes 50 epochs with 32 image minibatch size. The loss function is designed to minimize the average mean squared error between the ground truth current densities and output images of the inference process. Total training time is 11 hours using the TensorFlow 2.10.0 machine learning framework \cite{abadi2016tensorflow} and an rtx-a6000 graphics processing unit with 48 GB memory capacity.}
\end{table}

\subsection{Experimental Data} 

To investigate the utility of the MAGIC-UNet for experimental applications, we used QDM magnetic imaging data, both from a recently published study \cite{oliver_vector_2021} (by our group and collaborators) of a custom printed circuit board (PCB); and new data, acquired with the same QDM and PCB at variable standoff distances and noise levels. QDM magnetic images are obtained using a 4$\times$4$\,$mm ensemble NV diamond with a 1.7$\,\mu$m NV surface layer. (Details about the QDM are described in \cite{oliver_vector_2021}.)  A 5$\,$mA current is applied to the custom PCB to produce \~5$\,\mu$T Oersted magnetic fields measured by the NV centers. Magnetic field images are collected for three different current geometries, with standoff distances ranging from 36 to 65$\,\mu$m for small standoff and 486 to 650$\,\mu$m for large standoff. Larger standoff distances employ a translation stage to separate the diamond from the PCB.  Both single-shot and averaged measurements (20 or 400 exposures) are obtained, generating data with SNR between 4 and 87 for small standoff, and between 2 and 10 for large standoff.  The resolution of the experiment is limited by resolution of the camera, with a pixel size of about 2$\,\mu$m in the raw data; after cropping and binning to a 64x64 image each pixel is approximately 31$\,\mu$m. See the Supplemental Information for more information about the experimental process.

\subsection{Fourier Method Optimization}
\label{Fourier}
Fourier space transformations described by  \cite{roth_biomagnetism_2023}, \cite{broadway_improved_2020} are implemented to produce synthetic magnetic field data $B_x \text{, } B_y \text{ and } B_z$ from randomly generated, planar-current densities $J_x \text{ and } J_y$ at variable standoff distances. These standoff distances form a normal distribution with mean 50$\,\mu$m and standard deviation of 10$\,\mu$m, which approximates the experimental range of separation between the NV sensing layer and current traces in the PCB (see discussion in the Supplemental Information).

In order to show the utility of the MAGIC-UNet method, we implement careful optimization of the Fourier Method. $B_x$ and $B_y$ field components are used rather than $B_z$ for a more robust inversion technique, as described in \cite{broadway_improved_2020}.  In addition, Gaussian filter processing on magnetic field images is necessary to prevent errors from aliasing effects and a finite Fourier bandwidth. A Hann window function improves the resilience against noise of the Fourier inversion \cite{roth_using_1989}.  With these filters implemented, a 3D parameter optimization search is performed to find optimal Hann window and Gaussian filter strengths for multiple standoff distances, maximizing SSIM for synthetic data with respect to ground truth current densities.

\section{Data Availability}

Trained models, validation and experimental test data, and associated scripts are available at: \url{https://doi.org/10.7910/DVN/SD6PVP} \cite{Reed_Replication_2025}. Full size datasets for training additional models can be found at \url{https://doi.org/10.7910/DVN/SJDS2O} and \url{https://doi.org/10.7910/DVN/QPCS0I}.

\section{Acknowledgements}

We thank Bo Zhu and Edlyn Levine for their helpful contributions to the early development of this project; and Stephen DeVience for suggesting the acronym MAGIC-UNet. M.S.R. acknowledges the generous support of the Kiyomi and Ed Baird MGH Research Scholar award. This material is based on work by N.R.R. and D.B. supported by the National Science Foundation Graduate Research Fellowship Program under Grant Nos. DGE 2236417 and DGE 1840990. R.L.W. acknowledges support from the Laboratory for Physical Sciences Jumping Electron Quantum Fellowship Program under Award No. H9823022C0029; the IC Postdoctoral Fellowship under Award No. STEMWD00851; the U.S. Air Force Office of Scientific Research under Grant No. FA9550-22-1-0312, and the University of Maryland Quantum Technology Center. S.M.O. acknowledges support from the MITRE Independent Research and Development Program. K. S. O. acknowledges support through an appointment to the Intelligence Community Postdoctoral Research Fellowship Program at the University of Maryland, administered by Oak Ridge Institute for Science and Education through an interagency agreement between the U.S. Department of Energy and the Office of the Director of National Intelligence. Any opinions, findings, and conclusions or recommendations expressed in this material are those of the authors and do not necessarily reflect the views of the funding agencies.

Approved for Public Release; Distribution Unlimited. Public Release Case Number 24-1979. Portions of this technical data were produced for the U. S. Government under Contract No. FA8702-19-C-0001 and W56KGU-18-D-0004, and is subject to the Rights in Technical Data-Noncommercial Items Clause DFARS 252.227-7013 (FEB 2014). \copyright  2024 The MITRE Corporation.

\section{Author Contribution Statement}

*Denotes equal contribution between N.R.R. and D.B.. 

M.J.T. conceived the project and developed the early ideas with M.J.H.K. and D.B.. D.B. designed the MAGIC-UNet architecture and performed network optimization. D.B., N.R.R., and N.L.. worked on the machine learning code base and optimization. N.R.R. performed data generation and network training, optimized hyperparameters and characteristics of training data, managed and modified the project's central code base, coordinated other researchers on the creation of experimental and synthetic data, and developed and carried out the data analysis. S.M.O. and M.J.T. built the experimental setups and collected the experimental data, with assistance from K.S.O.. M.J.T., N.R.R., J.T., and D.M.D. created the synthetic data used for training and validation. D.M.D. implemented the Fourier Method analysis and optimized parameters for its performance. N.R.R. and D.M.D. wrote the manuscript, created data visualization tools, and generated the figures. R.L.W., M.S.R., M.J.H.K., and K.S.O. supervised the project. All authors reviewed the manuscript and were able to contribute to its editing and final content. 

\clearpage
\bibliography{main}

\begin{thebibliography}{61}%
\makeatletter
\providecommand \@ifxundefined [1]{%
 \@ifx{#1\undefined}
}%
\providecommand \@ifnum [1]{%
 \ifnum #1\expandafter \@firstoftwo
 \else \expandafter \@secondoftwo
 \fi
}%
\providecommand \@ifx [1]{%
 \ifx #1\expandafter \@firstoftwo
 \else \expandafter \@secondoftwo
 \fi
}%
\providecommand \natexlab [1]{#1}%
\providecommand \enquote  [1]{``#1''}%
\providecommand \bibnamefont  [1]{#1}%
\providecommand \bibfnamefont [1]{#1}%
\providecommand \citenamefont [1]{#1}%
\providecommand \href@noop [0]{\@secondoftwo}%
\providecommand \href [0]{\begingroup \@sanitize@url \@href}%
\providecommand \@href[1]{\@@startlink{#1}\@@href}%
\providecommand \@@href[1]{\endgroup#1\@@endlink}%
\providecommand \@sanitize@url [0]{\catcode `\\12\catcode `\$12\catcode `\&12\catcode `\#12\catcode `\^12\catcode `\_12\catcode `\%12\relax}%
\providecommand \@@startlink[1]{}%
\providecommand \@@endlink[0]{}%
\providecommand \url  [0]{\begingroup\@sanitize@url \@url }%
\providecommand \@url [1]{\endgroup\@href {#1}{\urlprefix }}%
\providecommand \urlprefix  [0]{URL }%
\providecommand \Eprint [0]{\href }%
\providecommand \doibase [0]{https://doi.org/}%
\providecommand \selectlanguage [0]{\@gobble}%
\providecommand \bibinfo  [0]{\@secondoftwo}%
\providecommand \bibfield  [0]{\@secondoftwo}%
\providecommand \translation [1]{[#1]}%
\providecommand \BibitemOpen [0]{}%
\providecommand \bibitemStop [0]{}%
\providecommand \bibitemNoStop [0]{.\EOS\space}%
\providecommand \EOS [0]{\spacefactor3000\relax}%
\providecommand \BibitemShut  [1]{\csname bibitem#1\endcsname}%
\let\auto@bib@innerbib\@empty
\bibitem [{\citenamefont {Orozco}(2009)}]{orozco_magnetic_2009}%
  \BibitemOpen
  \bibfield  {author} {\bibinfo {author} {\bibfnamefont {A.}~\bibnamefont {Orozco}},\ }\href {https://api.semanticscholar.org/CorpusID:251459118} {\bibfield  {journal} {\bibinfo  {journal} {EDFA Technical Articles}\ } (\bibinfo {year} {2009})}\BibitemShut {NoStop}%
\bibitem [{\citenamefont {Nowodzinski}\ \emph {et~al.}(2015)\citenamefont {Nowodzinski}, \citenamefont {Chipaux}, \citenamefont {Toraille}, \citenamefont {Jacques}, \citenamefont {Roch},\ and\ \citenamefont {Debuisschert}}]{nowodzinski_nitrogen-vacancy_2015}%
  \BibitemOpen
  \bibfield  {author} {\bibinfo {author} {\bibfnamefont {A.}~\bibnamefont {Nowodzinski}}, \bibinfo {author} {\bibfnamefont {M.}~\bibnamefont {Chipaux}}, \bibinfo {author} {\bibfnamefont {L.}~\bibnamefont {Toraille}}, \bibinfo {author} {\bibfnamefont {V.}~\bibnamefont {Jacques}}, \bibinfo {author} {\bibfnamefont {J.-F.}\ \bibnamefont {Roch}},\ and\ \bibinfo {author} {\bibfnamefont {T.}~\bibnamefont {Debuisschert}},\ }\href {https://doi.org/10.1016/j.microrel.2015.06.069} {\bibfield  {journal} {\bibinfo  {journal} {Microelectronics Reliability}\ }\textbf {\bibinfo {volume} {55}},\ \bibinfo {pages} {1549} (\bibinfo {year} {2015})}\BibitemShut {NoStop}%
\bibitem [{\citenamefont {Infante}\ \emph {et~al.}(2009{\natexlab{a}})\citenamefont {Infante}, \citenamefont {Perdu}, \citenamefont {Petremont},\ and\ \citenamefont {Lewis}}]{infante_new_2009}%
  \BibitemOpen
  \bibfield  {author} {\bibinfo {author} {\bibfnamefont {F.}~\bibnamefont {Infante}}, \bibinfo {author} {\bibfnamefont {P.}~\bibnamefont {Perdu}}, \bibinfo {author} {\bibfnamefont {S.}~\bibnamefont {Petremont}},\ and\ \bibinfo {author} {\bibfnamefont {D.}~\bibnamefont {Lewis}},\ }in\ \href {https://doi.org/10.1109/IPFA.2009.5232668} {\emph {\bibinfo {booktitle} {2009 16th {IEEE} {International} {Symposium} on the {Physical} and {Failure} {Analysis} of {Integrated} {Circuits}}}}\ (\bibinfo {year} {2009})\ pp.\ \bibinfo {pages} {208--212},\ \bibinfo {note} {iSSN: 1946-1550}\BibitemShut {NoStop}%
\bibitem [{lev(2020)}]{levine_backside_2020}%
  \BibitemOpen
  \href@noop {} {\emph {\bibinfo {title} {Backside Integrated Circuit Magnetic Field Imaging with a Quantum Diamond Microscope}}},\ International Symposium for Testing and Failure Analysis\ (\bibinfo {organization} {ASM International},\ \bibinfo {year} {2020})\BibitemShut {NoStop}%
\bibitem [{\citenamefont {Kehayias}\ \emph {et~al.}(2022)\citenamefont {Kehayias}, \citenamefont {Levine}, \citenamefont {Basso}, \citenamefont {Henshaw}, \citenamefont {Saleh~Ziabari}, \citenamefont {Titze}, \citenamefont {Haltli}, \citenamefont {Okoro}, \citenamefont {Tibbetts}, \citenamefont {Udoni}, \citenamefont {Bielejec}, \citenamefont {Lilly}, \citenamefont {Lu}, \citenamefont {Schwindt},\ and\ \citenamefont {Mounce}}]{kehayias_measurement_2022}%
  \BibitemOpen
  \bibfield  {author} {\bibinfo {author} {\bibfnamefont {P.}~\bibnamefont {Kehayias}}, \bibinfo {author} {\bibfnamefont {E.~V.}\ \bibnamefont {Levine}}, \bibinfo {author} {\bibfnamefont {L.}~\bibnamefont {Basso}}, \bibinfo {author} {\bibfnamefont {J.}~\bibnamefont {Henshaw}}, \bibinfo {author} {\bibfnamefont {M.}~\bibnamefont {Saleh~Ziabari}}, \bibinfo {author} {\bibfnamefont {M.}~\bibnamefont {Titze}}, \bibinfo {author} {\bibfnamefont {R.}~\bibnamefont {Haltli}}, \bibinfo {author} {\bibfnamefont {J.}~\bibnamefont {Okoro}}, \bibinfo {author} {\bibfnamefont {D.~R.}\ \bibnamefont {Tibbetts}}, \bibinfo {author} {\bibfnamefont {D.~M.}\ \bibnamefont {Udoni}}, \bibinfo {author} {\bibfnamefont {E.}~\bibnamefont {Bielejec}}, \bibinfo {author} {\bibfnamefont {M.~P.}\ \bibnamefont {Lilly}}, \bibinfo {author} {\bibfnamefont {T.-M.}\ \bibnamefont {Lu}}, \bibinfo {author} {\bibfnamefont {P.~D.~D.}\ \bibnamefont {Schwindt}},\ and\ \bibinfo {author} {\bibfnamefont {A.~M.}\ \bibnamefont {Mounce}},\ }\href
  {https://doi.org/10.1103/PhysRevApplied.17.014021} {\bibfield  {journal} {\bibinfo  {journal} {Physical Review Applied}\ }\textbf {\bibinfo {volume} {17}},\ \bibinfo {pages} {014021} (\bibinfo {year} {2022})}\BibitemShut {NoStop}%
\bibitem [{\citenamefont {Turner}(2020)}]{turner_quantum_2020}%
  \BibitemOpen
  \bibfield  {author} {\bibinfo {author} {\bibfnamefont {M.~J.}\ \bibnamefont {Turner}},\ }\emph {\bibinfo {title} {Quantum Diamond Microscopes for Biological Systems and Integrated Circuits}},\ \href@noop {} {Ph.D. thesis},\ \bibinfo  {school} {Harvard University} (\bibinfo {year} {2020})\BibitemShut {NoStop}%
\bibitem [{\citenamefont {Turner}\ \emph {et~al.}(2020)\citenamefont {Turner}, \citenamefont {Langellier}, \citenamefont {Bainbridge}, \citenamefont {Walters}, \citenamefont {Meesala}, \citenamefont {Babinec}, \citenamefont {Kehayias}, \citenamefont {Yacoby}, \citenamefont {Hu}, \citenamefont {Lon{\v{c}}ar} \emph {et~al.}}]{turner_magnetic_2020}%
  \BibitemOpen
  \bibfield  {author} {\bibinfo {author} {\bibfnamefont {M.~J.}\ \bibnamefont {Turner}}, \bibinfo {author} {\bibfnamefont {N.}~\bibnamefont {Langellier}}, \bibinfo {author} {\bibfnamefont {R.}~\bibnamefont {Bainbridge}}, \bibinfo {author} {\bibfnamefont {D.}~\bibnamefont {Walters}}, \bibinfo {author} {\bibfnamefont {S.}~\bibnamefont {Meesala}}, \bibinfo {author} {\bibfnamefont {T.~M.}\ \bibnamefont {Babinec}}, \bibinfo {author} {\bibfnamefont {P.}~\bibnamefont {Kehayias}}, \bibinfo {author} {\bibfnamefont {A.}~\bibnamefont {Yacoby}}, \bibinfo {author} {\bibfnamefont {E.}~\bibnamefont {Hu}}, \bibinfo {author} {\bibfnamefont {M.}~\bibnamefont {Lon{\v{c}}ar}}, \emph {et~al.},\ }\href {www.doi-org.com/10.1103/PhysRevApplied.14.014097} {\bibfield  {journal} {\bibinfo  {journal} {Physical Review Applied}\ }\textbf {\bibinfo {volume} {14}},\ \bibinfo {pages} {014097} (\bibinfo {year} {2020})}\BibitemShut {NoStop}%
\bibitem [{\citenamefont {Infante}\ \emph {et~al.}(2009{\natexlab{b}})\citenamefont {Infante}, \citenamefont {Perdu},\ and\ \citenamefont {Lewis}}]{infante_magnetic_2009}%
  \BibitemOpen
  \bibfield  {author} {\bibinfo {author} {\bibfnamefont {F.}~\bibnamefont {Infante}}, \bibinfo {author} {\bibfnamefont {P.}~\bibnamefont {Perdu}},\ and\ \bibinfo {author} {\bibfnamefont {D.}~\bibnamefont {Lewis}},\ }\href {https://doi.org/10.1016/j.microrel.2009.06.041} {\bibfield  {journal} {\bibinfo  {journal} {Microelectronics Reliability}\ }\bibinfo {series} {20th {European} {Symposium} on the {Reliability} of {Electron} {Devices}, {Failure} {Physics} and {Analysis}},\ \textbf {\bibinfo {volume} {49}},\ \bibinfo {pages} {1169} (\bibinfo {year} {2009}{\natexlab{b}})}\BibitemShut {NoStop}%
\bibitem [{oli(2021)}]{oliver_vector_2021}%
  \BibitemOpen
  \href {https://doi.org/10.31399/asm.cp.istfa2021p0096} {\emph {\bibinfo {title} {47th International Symposium for Testing and Failure Analysis}}},\ International Symposium for Testing and Failure Analysis\ (\bibinfo {organization} {ASM},\ \bibinfo {year} {2021})\BibitemShut {NoStop}%
\bibitem [{fel(2007)}]{felt_construction_2007}%
  \BibitemOpen
  \href {https://doi.org/10.31399/asm.cp.istfa2007p0197} {\emph {\bibinfo {title} {Construction of a 3-{D} {Current} {Path} {Using} {Magnetic} {Current} {Imaging}}}},\ 33rd International Symposium for Testing and Failure Analysis\ (\bibinfo  {publisher} {ASM International},\ \bibinfo {year} {2007})\BibitemShut {NoStop}%
\bibitem [{\citenamefont {Hu}\ \emph {et~al.}(2019)\citenamefont {Hu}, \citenamefont {Iwata}, \citenamefont {Mohammadi}, \citenamefont {Silletta}, \citenamefont {Wickenbrock}, \citenamefont {Blanchard}, \citenamefont {Budker},\ and\ \citenamefont {Jerschow}}]{hu_battery_2019}%
  \BibitemOpen
  \bibfield  {author} {\bibinfo {author} {\bibfnamefont {Y.}~\bibnamefont {Hu}}, \bibinfo {author} {\bibfnamefont {G.~Z.}\ \bibnamefont {Iwata}}, \bibinfo {author} {\bibfnamefont {M.}~\bibnamefont {Mohammadi}}, \bibinfo {author} {\bibfnamefont {E.~V.}\ \bibnamefont {Silletta}}, \bibinfo {author} {\bibfnamefont {A.}~\bibnamefont {Wickenbrock}}, \bibinfo {author} {\bibfnamefont {J.~W.}\ \bibnamefont {Blanchard}}, \bibinfo {author} {\bibfnamefont {D.}~\bibnamefont {Budker}},\ and\ \bibinfo {author} {\bibfnamefont {A.}~\bibnamefont {Jerschow}},\ }\href {https://doi.org/10.48550/arXiv.1905.12507} {\bibinfo {title} {Battery {Diagnostics} with {Sensitive} {Magnetometry}}} (\bibinfo {year} {2019}),\ \bibinfo {note} {arXiv:1905.12507 [physics]}\BibitemShut {NoStop}%
\bibitem [{\citenamefont {Hu}\ \emph {et~al.}(2020)\citenamefont {Hu}, \citenamefont {Iwata}, \citenamefont {Mohammadi}, \citenamefont {Silletta}, \citenamefont {Wickenbrock}, \citenamefont {Blanchard}, \citenamefont {Budker},\ and\ \citenamefont {Jerschow}}]{hu_sensitive_2020}%
  \BibitemOpen
  \bibfield  {author} {\bibinfo {author} {\bibfnamefont {Y.}~\bibnamefont {Hu}}, \bibinfo {author} {\bibfnamefont {G.~Z.}\ \bibnamefont {Iwata}}, \bibinfo {author} {\bibfnamefont {M.}~\bibnamefont {Mohammadi}}, \bibinfo {author} {\bibfnamefont {E.~V.}\ \bibnamefont {Silletta}}, \bibinfo {author} {\bibfnamefont {A.}~\bibnamefont {Wickenbrock}}, \bibinfo {author} {\bibfnamefont {J.~W.}\ \bibnamefont {Blanchard}}, \bibinfo {author} {\bibfnamefont {D.}~\bibnamefont {Budker}},\ and\ \bibinfo {author} {\bibfnamefont {A.}~\bibnamefont {Jerschow}},\ }\href {https://doi.org/10.1073/pnas.1917172117} {\bibfield  {journal} {\bibinfo  {journal} {Proceedings of the National Academy of Sciences}\ }\textbf {\bibinfo {volume} {117}},\ \bibinfo {pages} {10667} (\bibinfo {year} {2020})},\ \bibinfo {note} {publisher: Proceedings of the National Academy of Sciences}\BibitemShut {NoStop}%
\bibitem [{\citenamefont {Bason}\ \emph {et~al.}(2022)\citenamefont {Bason}, \citenamefont {Coussens}, \citenamefont {Withers}, \citenamefont {Abel}, \citenamefont {Kendall},\ and\ \citenamefont {Krüger}}]{bason_non-invasive_2022}%
  \BibitemOpen
  \bibfield  {author} {\bibinfo {author} {\bibfnamefont {M.~G.}\ \bibnamefont {Bason}}, \bibinfo {author} {\bibfnamefont {T.}~\bibnamefont {Coussens}}, \bibinfo {author} {\bibfnamefont {M.}~\bibnamefont {Withers}}, \bibinfo {author} {\bibfnamefont {C.}~\bibnamefont {Abel}}, \bibinfo {author} {\bibfnamefont {G.}~\bibnamefont {Kendall}},\ and\ \bibinfo {author} {\bibfnamefont {P.}~\bibnamefont {Krüger}},\ }\href {https://doi.org/10.1016/j.jpowsour.2022.231312} {\bibfield  {journal} {\bibinfo  {journal} {Journal of Power Sources}\ }\textbf {\bibinfo {volume} {533}},\ \bibinfo {pages} {231312} (\bibinfo {year} {2022})}\BibitemShut {NoStop}%
\bibitem [{\citenamefont {Brauchle}\ \emph {et~al.}(2021)\citenamefont {Brauchle}, \citenamefont {Grimsmann}, \citenamefont {von Kessel},\ and\ \citenamefont {Birke}}]{brauchle_direct_2021}%
  \BibitemOpen
  \bibfield  {author} {\bibinfo {author} {\bibfnamefont {F.}~\bibnamefont {Brauchle}}, \bibinfo {author} {\bibfnamefont {F.}~\bibnamefont {Grimsmann}}, \bibinfo {author} {\bibfnamefont {O.}~\bibnamefont {von Kessel}},\ and\ \bibinfo {author} {\bibfnamefont {K.~P.}\ \bibnamefont {Birke}},\ }\href {https://doi.org/10.1016/j.jpowsour.2021.230292} {\bibfield  {journal} {\bibinfo  {journal} {Journal of Power Sources}\ }\textbf {\bibinfo {volume} {507}},\ \bibinfo {pages} {230292} (\bibinfo {year} {2021})}\BibitemShut {NoStop}%
\bibitem [{\citenamefont {Scholten}\ \emph {et~al.}(2022)\citenamefont {Scholten}, \citenamefont {Abrahams}, \citenamefont {Johnson}, \citenamefont {Healey}, \citenamefont {Robertson}, \citenamefont {Simpson}, \citenamefont {Stacey}, \citenamefont {Onoda}, \citenamefont {Ohshima}, \citenamefont {Kho}, \citenamefont {Ibarra~Michel}, \citenamefont {Bullock}, \citenamefont {Hollenberg},\ and\ \citenamefont {Tetienne}}]{scholten_imaging_2022}%
  \BibitemOpen
  \bibfield  {author} {\bibinfo {author} {\bibfnamefont {S.}~\bibnamefont {Scholten}}, \bibinfo {author} {\bibfnamefont {G.}~\bibnamefont {Abrahams}}, \bibinfo {author} {\bibfnamefont {B.}~\bibnamefont {Johnson}}, \bibinfo {author} {\bibfnamefont {A.}~\bibnamefont {Healey}}, \bibinfo {author} {\bibfnamefont {I.}~\bibnamefont {Robertson}}, \bibinfo {author} {\bibfnamefont {D.}~\bibnamefont {Simpson}}, \bibinfo {author} {\bibfnamefont {A.}~\bibnamefont {Stacey}}, \bibinfo {author} {\bibfnamefont {S.}~\bibnamefont {Onoda}}, \bibinfo {author} {\bibfnamefont {T.}~\bibnamefont {Ohshima}}, \bibinfo {author} {\bibfnamefont {T.}~\bibnamefont {Kho}}, \bibinfo {author} {\bibfnamefont {J.}~\bibnamefont {Ibarra~Michel}}, \bibinfo {author} {\bibfnamefont {J.}~\bibnamefont {Bullock}}, \bibinfo {author} {\bibfnamefont {L.}~\bibnamefont {Hollenberg}},\ and\ \bibinfo {author} {\bibfnamefont {J.-P.}\ \bibnamefont {Tetienne}},\ }\href {https://doi.org/10.1103/PhysRevApplied.18.014041} {\bibfield  {journal} {\bibinfo  {journal}
  {Physical Review Applied}\ }\textbf {\bibinfo {volume} {18}},\ \bibinfo {pages} {014041} (\bibinfo {year} {2022})},\ \bibinfo {note} {publisher: American Physical Society}\BibitemShut {NoStop}%
\bibitem [{\citenamefont {Kaufmann}\ \emph {et~al.}(2021)\citenamefont {Kaufmann}, \citenamefont {Lausch}, \citenamefont {Lin}, \citenamefont {Rudolph}, \citenamefont {Hahn},\ and\ \citenamefont {Patzold}}]{kaufmann_evaluation_2021}%
  \BibitemOpen
  \bibfield  {author} {\bibinfo {author} {\bibfnamefont {K.}~\bibnamefont {Kaufmann}}, \bibinfo {author} {\bibfnamefont {D.}~\bibnamefont {Lausch}}, \bibinfo {author} {\bibfnamefont {C.-M.}\ \bibnamefont {Lin}}, \bibinfo {author} {\bibfnamefont {M.}~\bibnamefont {Rudolph}}, \bibinfo {author} {\bibfnamefont {D.}~\bibnamefont {Hahn}},\ and\ \bibinfo {author} {\bibfnamefont {M.}~\bibnamefont {Patzold}},\ }\href {https://onlinelibrary.wiley.com/doi/abs/10.1002/pssa.202000292} {\bibfield  {journal} {\bibinfo  {journal} {physica status solidi (a)}\ }\textbf {\bibinfo {volume} {218}},\ \bibinfo {pages} {2000292} (\bibinfo {year} {2021})}\BibitemShut {NoStop}%
\bibitem [{\citenamefont {Chikumoto}\ \emph {et~al.}(2021)\citenamefont {Chikumoto}, \citenamefont {Yamaguchi},\ and\ \citenamefont {Shyshkin}}]{chikumoto_characterization_2021}%
  \BibitemOpen
  \bibfield  {author} {\bibinfo {author} {\bibfnamefont {N.}~\bibnamefont {Chikumoto}}, \bibinfo {author} {\bibfnamefont {S.}~\bibnamefont {Yamaguchi}},\ and\ \bibinfo {author} {\bibfnamefont {O.}~\bibnamefont {Shyshkin}},\ }\href {https://doi.org/10.1109/TASC.2021.3070126} {\bibfield  {journal} {\bibinfo  {journal} {IEEE Transactions on Applied Superconductivity}\ }\textbf {\bibinfo {volume} {31}},\ \bibinfo {pages} {1} (\bibinfo {year} {2021})},\ \bibinfo {note} {conference Name: IEEE Transactions on Applied Superconductivity}\BibitemShut {NoStop}%
\bibitem [{\citenamefont {Sakai}\ \emph {et~al.}(2023)\citenamefont {Sakai}, \citenamefont {Yamada}, \citenamefont {Zhao}, \citenamefont {Zhu},\ and\ \citenamefont {Inoue}}]{sakai_effect_2023}%
  \BibitemOpen
  \bibfield  {author} {\bibinfo {author} {\bibfnamefont {S.}~\bibnamefont {Sakai}}, \bibinfo {author} {\bibfnamefont {Y.}~\bibnamefont {Yamada}}, \bibinfo {author} {\bibfnamefont {Y.}~\bibnamefont {Zhao}}, \bibinfo {author} {\bibfnamefont {J.}~\bibnamefont {Zhu}},\ and\ \bibinfo {author} {\bibfnamefont {M.}~\bibnamefont {Inoue}},\ }\href {https://doi.org/10.1109/TASC.2023.3249644} {\bibfield  {journal} {\bibinfo  {journal} {IEEE Transactions on Applied Superconductivity}\ }\textbf {\bibinfo {volume} {33}},\ \bibinfo {pages} {1} (\bibinfo {year} {2023})},\ \bibinfo {note} {conference Name: IEEE Transactions on Applied Superconductivity}\BibitemShut {NoStop}%
\bibitem [{\citenamefont {Bevington}\ \emph {et~al.}(2018)\citenamefont {Bevington}, \citenamefont {Gartman}, \citenamefont {Chalupczak}, \citenamefont {Deans}, \citenamefont {Marmugi},\ and\ \citenamefont {Renzoni}}]{bevington_non-destructive_2018}%
  \BibitemOpen
  \bibfield  {author} {\bibinfo {author} {\bibfnamefont {P.}~\bibnamefont {Bevington}}, \bibinfo {author} {\bibfnamefont {R.}~\bibnamefont {Gartman}}, \bibinfo {author} {\bibfnamefont {W.}~\bibnamefont {Chalupczak}}, \bibinfo {author} {\bibfnamefont {C.}~\bibnamefont {Deans}}, \bibinfo {author} {\bibfnamefont {L.}~\bibnamefont {Marmugi}},\ and\ \bibinfo {author} {\bibfnamefont {F.}~\bibnamefont {Renzoni}},\ }\href {https://doi.org/10.1063/1.5042033} {\bibfield  {journal} {\bibinfo  {journal} {Applied Physics Letters}\ }\textbf {\bibinfo {volume} {113}},\ \bibinfo {pages} {063503} (\bibinfo {year} {2018})}\BibitemShut {NoStop}%
\bibitem [{\citenamefont {Wikswo}(1996)}]{wikswo_magnetic_1996}%
  \BibitemOpen
  \bibfield  {author} {\bibinfo {author} {\bibfnamefont {J.~P.}\ \bibnamefont {Wikswo}},\ }in\ \href {https://doi.org/10.1007/978-94-011-5674-5_16} {\emph {\bibinfo {booktitle} {{SQUID} {Sensors}: {Fundamentals}, {Fabrication} and {Applications}}}},\ \bibinfo {series and number} {{NATO} {ASI} {Series}},\ \bibinfo {editor} {edited by\ \bibinfo {editor} {\bibfnamefont {H.}~\bibnamefont {Weinstock}}}\ (\bibinfo  {publisher} {Springer Netherlands},\ \bibinfo {address} {Dordrecht},\ \bibinfo {year} {1996})\ pp.\ \bibinfo {pages} {629--695}\BibitemShut {NoStop}%
\bibitem [{\citenamefont {Wu}\ \emph {et~al.}(2021)\citenamefont {Wu}, \citenamefont {Higashikawa}, \citenamefont {Imamura}, \citenamefont {Xu}, \citenamefont {Ma},\ and\ \citenamefont {Kiss}}]{wu_characterization_2021}%
  \BibitemOpen
  \bibfield  {author} {\bibinfo {author} {\bibfnamefont {Z.}~\bibnamefont {Wu}}, \bibinfo {author} {\bibfnamefont {K.}~\bibnamefont {Higashikawa}}, \bibinfo {author} {\bibfnamefont {K.}~\bibnamefont {Imamura}}, \bibinfo {author} {\bibfnamefont {Z.}~\bibnamefont {Xu}}, \bibinfo {author} {\bibfnamefont {Y.}~\bibnamefont {Ma}},\ and\ \bibinfo {author} {\bibfnamefont {T.}~\bibnamefont {Kiss}},\ }\href {https://doi.org/10.1109/TASC.2021.3059988} {\bibfield  {journal} {\bibinfo  {journal} {IEEE Transactions on Applied Superconductivity}\ }\textbf {\bibinfo {volume} {31}},\ \bibinfo {pages} {1} (\bibinfo {year} {2021})},\ \bibinfo {note} {conference Name: IEEE Transactions on Applied Superconductivity}\BibitemShut {NoStop}%
\bibitem [{\citenamefont {Kirtley}(2010)}]{kirtley_fundamental_2010}%
  \BibitemOpen
  \bibfield  {author} {\bibinfo {author} {\bibfnamefont {J.~R.}\ \bibnamefont {Kirtley}},\ }\href {https://doi.org/10.1088/0034-4885/73/12/126501} {\bibfield  {journal} {\bibinfo  {journal} {Reports on Progress in Physics}\ }\textbf {\bibinfo {volume} {73}},\ \bibinfo {pages} {126501} (\bibinfo {year} {2010})}\BibitemShut {NoStop}%
\bibitem [{\citenamefont {Marchiori}\ \emph {et~al.}(2022)\citenamefont {Marchiori}, \citenamefont {Ceccarelli}, \citenamefont {Rossi}, \citenamefont {Romagnoli}, \citenamefont {Herrmann}, \citenamefont {Besse}, \citenamefont {Krinner}, \citenamefont {Wallraff},\ and\ \citenamefont {Poggio}}]{marchiori_magnetic_2022}%
  \BibitemOpen
  \bibfield  {author} {\bibinfo {author} {\bibfnamefont {E.}~\bibnamefont {Marchiori}}, \bibinfo {author} {\bibfnamefont {L.}~\bibnamefont {Ceccarelli}}, \bibinfo {author} {\bibfnamefont {N.}~\bibnamefont {Rossi}}, \bibinfo {author} {\bibfnamefont {G.}~\bibnamefont {Romagnoli}}, \bibinfo {author} {\bibfnamefont {J.}~\bibnamefont {Herrmann}}, \bibinfo {author} {\bibfnamefont {J.-C.}\ \bibnamefont {Besse}}, \bibinfo {author} {\bibfnamefont {S.}~\bibnamefont {Krinner}}, \bibinfo {author} {\bibfnamefont {A.}~\bibnamefont {Wallraff}},\ and\ \bibinfo {author} {\bibfnamefont {M.}~\bibnamefont {Poggio}},\ }\href {https://doi.org/10.1063/5.0103597} {\bibfield  {journal} {\bibinfo  {journal} {Applied Physics Letters}\ }\textbf {\bibinfo {volume} {121}},\ \bibinfo {pages} {052601} (\bibinfo {year} {2022})}\BibitemShut {NoStop}%
\bibitem [{\citenamefont {Ku}\ \emph {et~al.}(2020)\citenamefont {Ku}, \citenamefont {Zhou}, \citenamefont {Li}, \citenamefont {Shin}, \citenamefont {Shi}, \citenamefont {Burch}, \citenamefont {Anderson}, \citenamefont {Pierce}, \citenamefont {Xie}, \citenamefont {Hamo}, \citenamefont {Vool}, \citenamefont {Zhang}, \citenamefont {Casola}, \citenamefont {Taniguchi}, \citenamefont {Watanabe}, \citenamefont {Fogler}, \citenamefont {Kim}, \citenamefont {Yacoby},\ and\ \citenamefont {Walsworth}}]{ku_imaging_2020}%
  \BibitemOpen
  \bibfield  {author} {\bibinfo {author} {\bibfnamefont {M.~J.~H.}\ \bibnamefont {Ku}}, \bibinfo {author} {\bibfnamefont {T.~X.}\ \bibnamefont {Zhou}}, \bibinfo {author} {\bibfnamefont {Q.}~\bibnamefont {Li}}, \bibinfo {author} {\bibfnamefont {Y.~J.}\ \bibnamefont {Shin}}, \bibinfo {author} {\bibfnamefont {J.~K.}\ \bibnamefont {Shi}}, \bibinfo {author} {\bibfnamefont {C.}~\bibnamefont {Burch}}, \bibinfo {author} {\bibfnamefont {L.~E.}\ \bibnamefont {Anderson}}, \bibinfo {author} {\bibfnamefont {A.~T.}\ \bibnamefont {Pierce}}, \bibinfo {author} {\bibfnamefont {Y.}~\bibnamefont {Xie}}, \bibinfo {author} {\bibfnamefont {A.}~\bibnamefont {Hamo}}, \bibinfo {author} {\bibfnamefont {U.}~\bibnamefont {Vool}}, \bibinfo {author} {\bibfnamefont {H.}~\bibnamefont {Zhang}}, \bibinfo {author} {\bibfnamefont {F.}~\bibnamefont {Casola}}, \bibinfo {author} {\bibfnamefont {T.}~\bibnamefont {Taniguchi}}, \bibinfo {author} {\bibfnamefont {K.}~\bibnamefont {Watanabe}}, \bibinfo {author} {\bibfnamefont {M.~M.}\ \bibnamefont
  {Fogler}}, \bibinfo {author} {\bibfnamefont {P.}~\bibnamefont {Kim}}, \bibinfo {author} {\bibfnamefont {A.}~\bibnamefont {Yacoby}},\ and\ \bibinfo {author} {\bibfnamefont {R.~L.}\ \bibnamefont {Walsworth}},\ }\href {https://doi.org/10.1038/s41586-020-2507-2} {\bibfield  {journal} {\bibinfo  {journal} {Nature}\ }\textbf {\bibinfo {volume} {583}},\ \bibinfo {pages} {537} (\bibinfo {year} {2020})}\BibitemShut {NoStop}%
\bibitem [{\citenamefont {Feng}\ \emph {et~al.}(2022)\citenamefont {Feng}, \citenamefont {Zhu}, \citenamefont {Lin}, \citenamefont {Lian}, \citenamefont {Wang}, \citenamefont {Li}, \citenamefont {Yao}, \citenamefont {He}, \citenamefont {Pan}, \citenamefont {Wu}, \citenamefont {Zhang}, \citenamefont {Wang}, \citenamefont {Zhou}, \citenamefont {Shen},\ and\ \citenamefont {Wang}}]{feng_helical_2022}%
  \BibitemOpen
  \bibfield  {author} {\bibinfo {author} {\bibfnamefont {Y.}~\bibnamefont {Feng}}, \bibinfo {author} {\bibfnamefont {J.}~\bibnamefont {Zhu}}, \bibinfo {author} {\bibfnamefont {W.}~\bibnamefont {Lin}}, \bibinfo {author} {\bibfnamefont {Z.}~\bibnamefont {Lian}}, \bibinfo {author} {\bibfnamefont {Y.}~\bibnamefont {Wang}}, \bibinfo {author} {\bibfnamefont {H.}~\bibnamefont {Li}}, \bibinfo {author} {\bibfnamefont {H.}~\bibnamefont {Yao}}, \bibinfo {author} {\bibfnamefont {Q.}~\bibnamefont {He}}, \bibinfo {author} {\bibfnamefont {Y.}~\bibnamefont {Pan}}, \bibinfo {author} {\bibfnamefont {Y.}~\bibnamefont {Wu}}, \bibinfo {author} {\bibfnamefont {J.}~\bibnamefont {Zhang}}, \bibinfo {author} {\bibfnamefont {Y.}~\bibnamefont {Wang}}, \bibinfo {author} {\bibfnamefont {X.}~\bibnamefont {Zhou}}, \bibinfo {author} {\bibfnamefont {J.}~\bibnamefont {Shen}},\ and\ \bibinfo {author} {\bibfnamefont {Y.}~\bibnamefont {Wang}},\ }\href {https://doi.org/10.1021/acs.nanolett.2c02701} {\bibfield  {journal} {\bibinfo  {journal} {Nano
  Letters}\ }\textbf {\bibinfo {volume} {22}},\ \bibinfo {pages} {7606} (\bibinfo {year} {2022})},\ \bibinfo {note} {publisher: American Chemical Society}\BibitemShut {NoStop}%
\bibitem [{\citenamefont {Palm}\ \emph {et~al.}(2022)\citenamefont {Palm}, \citenamefont {Huxter}, \citenamefont {Welter}, \citenamefont {Ernst}, \citenamefont {Scheidegger}, \citenamefont {Diesch}, \citenamefont {Chang}, \citenamefont {Rickhaus}, \citenamefont {Taniguchi}, \citenamefont {Watanabe}, \citenamefont {Ensslin},\ and\ \citenamefont {Degen}}]{palm_imaging_2022}%
  \BibitemOpen
  \bibfield  {author} {\bibinfo {author} {\bibfnamefont {M.}~\bibnamefont {Palm}}, \bibinfo {author} {\bibfnamefont {W.}~\bibnamefont {Huxter}}, \bibinfo {author} {\bibfnamefont {P.}~\bibnamefont {Welter}}, \bibinfo {author} {\bibfnamefont {S.}~\bibnamefont {Ernst}}, \bibinfo {author} {\bibfnamefont {P.}~\bibnamefont {Scheidegger}}, \bibinfo {author} {\bibfnamefont {S.}~\bibnamefont {Diesch}}, \bibinfo {author} {\bibfnamefont {K.}~\bibnamefont {Chang}}, \bibinfo {author} {\bibfnamefont {P.}~\bibnamefont {Rickhaus}}, \bibinfo {author} {\bibfnamefont {T.}~\bibnamefont {Taniguchi}}, \bibinfo {author} {\bibfnamefont {K.}~\bibnamefont {Watanabe}}, \bibinfo {author} {\bibfnamefont {K.}~\bibnamefont {Ensslin}},\ and\ \bibinfo {author} {\bibfnamefont {C.}~\bibnamefont {Degen}},\ }\href {https://doi.org/10.1103/PhysRevApplied.17.054008} {\bibfield  {journal} {\bibinfo  {journal} {Physical Review Applied}\ }\textbf {\bibinfo {volume} {17}},\ \bibinfo {pages} {054008} (\bibinfo {year} {2022})},\ \bibinfo {note}
  {publisher: American Physical Society}\BibitemShut {NoStop}%
\bibitem [{\citenamefont {Bjørlig}\ \emph {et~al.}(2022)\citenamefont {Bjørlig}, \citenamefont {Christensen}, \citenamefont {Erlandsen}, \citenamefont {Pryds},\ and\ \citenamefont {Kalisky}}]{bjorlig_current_2022}%
  \BibitemOpen
  \bibfield  {author} {\bibinfo {author} {\bibfnamefont {A.~V.}\ \bibnamefont {Bjørlig}}, \bibinfo {author} {\bibfnamefont {D.~V.}\ \bibnamefont {Christensen}}, \bibinfo {author} {\bibfnamefont {R.}~\bibnamefont {Erlandsen}}, \bibinfo {author} {\bibfnamefont {N.}~\bibnamefont {Pryds}},\ and\ \bibinfo {author} {\bibfnamefont {B.}~\bibnamefont {Kalisky}},\ }\href {https://doi.org/10.1021/acsaelm.2c00264} {\bibfield  {journal} {\bibinfo  {journal} {ACS Applied Electronic Materials}\ }\textbf {\bibinfo {volume} {4}},\ \bibinfo {pages} {3421} (\bibinfo {year} {2022})},\ \bibinfo {note} {publisher: American Chemical Society}\BibitemShut {NoStop}%
\bibitem [{\citenamefont {Basso}\ \emph {et~al.}(2023)\citenamefont {Basso}, \citenamefont {Kehayias}, \citenamefont {Henshaw}, \citenamefont {Saleh~Ziabari}, \citenamefont {Byeon}, \citenamefont {Lilly}, \citenamefont {Bussmann}, \citenamefont {Campbell}, \citenamefont {Misra},\ and\ \citenamefont {Mounce}}]{basso_electric_2023}%
  \BibitemOpen
  \bibfield  {author} {\bibinfo {author} {\bibfnamefont {L.}~\bibnamefont {Basso}}, \bibinfo {author} {\bibfnamefont {P.}~\bibnamefont {Kehayias}}, \bibinfo {author} {\bibfnamefont {J.}~\bibnamefont {Henshaw}}, \bibinfo {author} {\bibfnamefont {M.}~\bibnamefont {Saleh~Ziabari}}, \bibinfo {author} {\bibfnamefont {H.}~\bibnamefont {Byeon}}, \bibinfo {author} {\bibfnamefont {M.~P.}\ \bibnamefont {Lilly}}, \bibinfo {author} {\bibfnamefont {E.}~\bibnamefont {Bussmann}}, \bibinfo {author} {\bibfnamefont {D.~M.}\ \bibnamefont {Campbell}}, \bibinfo {author} {\bibfnamefont {S.}~\bibnamefont {Misra}},\ and\ \bibinfo {author} {\bibfnamefont {A.~M.}\ \bibnamefont {Mounce}},\ }\href {https://doi.org/10.1088/1361-6528/ac95a0} {\bibfield  {journal} {\bibinfo  {journal} {Nanotechnology}\ }\textbf {\bibinfo {volume} {34}},\ \bibinfo {pages} {015001} (\bibinfo {year} {2023})}\BibitemShut {NoStop}%
\bibitem [{\citenamefont {Lillie}\ \emph {et~al.}(2019)\citenamefont {Lillie}, \citenamefont {Dontschuk}, \citenamefont {Broadway}, \citenamefont {Creedon}, \citenamefont {Hollenberg},\ and\ \citenamefont {Tetienne}}]{lillie_imaging_2019}%
  \BibitemOpen
  \bibfield  {author} {\bibinfo {author} {\bibfnamefont {S.~E.}\ \bibnamefont {Lillie}}, \bibinfo {author} {\bibfnamefont {N.}~\bibnamefont {Dontschuk}}, \bibinfo {author} {\bibfnamefont {D.~A.}\ \bibnamefont {Broadway}}, \bibinfo {author} {\bibfnamefont {D.~L.}\ \bibnamefont {Creedon}}, \bibinfo {author} {\bibfnamefont {L.~C.}\ \bibnamefont {Hollenberg}},\ and\ \bibinfo {author} {\bibfnamefont {J.-P.}\ \bibnamefont {Tetienne}},\ }\href {https://doi.org/10.1103/PhysRevApplied.12.024018} {\bibfield  {journal} {\bibinfo  {journal} {Physical Review Applied}\ }\textbf {\bibinfo {volume} {12}},\ \bibinfo {pages} {024018} (\bibinfo {year} {2019})},\ \bibinfo {note} {publisher: American Physical Society}\BibitemShut {NoStop}%
\bibitem [{\citenamefont {Wickenbrock}\ \emph {et~al.}(2016)\citenamefont {Wickenbrock}, \citenamefont {Leefer}, \citenamefont {Blanchard},\ and\ \citenamefont {Budker}}]{wickenbrock_eddy_2016}%
  \BibitemOpen
  \bibfield  {author} {\bibinfo {author} {\bibfnamefont {A.}~\bibnamefont {Wickenbrock}}, \bibinfo {author} {\bibfnamefont {N.}~\bibnamefont {Leefer}}, \bibinfo {author} {\bibfnamefont {J.~W.}\ \bibnamefont {Blanchard}},\ and\ \bibinfo {author} {\bibfnamefont {D.}~\bibnamefont {Budker}},\ }\href {https://doi.org/10.1063/1.4948534} {\bibfield  {journal} {\bibinfo  {journal} {Applied Physics Letters}\ }\textbf {\bibinfo {volume} {108}},\ \bibinfo {pages} {183507} (\bibinfo {year} {2016})}\BibitemShut {NoStop}%
\bibitem [{\citenamefont {Roth}(2023)}]{roth_biomagnetism_2023}%
  \BibitemOpen
  \bibfield  {author} {\bibinfo {author} {\bibfnamefont {B.~J.}\ \bibnamefont {Roth}},\ }\href {https://doi.org/10.3390/s23094218} {\bibfield  {journal} {\bibinfo  {journal} {Sensors}\ }\textbf {\bibinfo {volume} {23}},\ \bibinfo {pages} {4218} (\bibinfo {year} {2023})}\BibitemShut {NoStop}%
\bibitem [{\citenamefont {Leder}\ \emph {et~al.}(1998)\citenamefont {Leder}, \citenamefont {Pohl}, \citenamefont {Michaelsen}, \citenamefont {Fritschi}, \citenamefont {Huck}, \citenamefont {Eichhorn}, \citenamefont {Müller},\ and\ \citenamefont {Nowak}}]{leder_noninvasive_1998}%
  \BibitemOpen
  \bibfield  {author} {\bibinfo {author} {\bibfnamefont {U.}~\bibnamefont {Leder}}, \bibinfo {author} {\bibfnamefont {H.-P.}\ \bibnamefont {Pohl}}, \bibinfo {author} {\bibfnamefont {S.}~\bibnamefont {Michaelsen}}, \bibinfo {author} {\bibfnamefont {T.}~\bibnamefont {Fritschi}}, \bibinfo {author} {\bibfnamefont {M.}~\bibnamefont {Huck}}, \bibinfo {author} {\bibfnamefont {J.}~\bibnamefont {Eichhorn}}, \bibinfo {author} {\bibfnamefont {S.}~\bibnamefont {Müller}},\ and\ \bibinfo {author} {\bibfnamefont {H.}~\bibnamefont {Nowak}},\ }\href {https://doi.org/10.1016/S0167-5273(97)00326-4} {\bibfield  {journal} {\bibinfo  {journal} {International Journal of Cardiology}\ }\textbf {\bibinfo {volume} {64}},\ \bibinfo {pages} {83} (\bibinfo {year} {1998})}\BibitemShut {NoStop}%
\bibitem [{\citenamefont {Merwa}\ \emph {et~al.}(2005)\citenamefont {Merwa}, \citenamefont {Hollaus}, \citenamefont {Brunner},\ and\ \citenamefont {Scharfetter}}]{merwa_solution_2005}%
  \BibitemOpen
  \bibfield  {author} {\bibinfo {author} {\bibfnamefont {R.}~\bibnamefont {Merwa}}, \bibinfo {author} {\bibfnamefont {K.}~\bibnamefont {Hollaus}}, \bibinfo {author} {\bibfnamefont {P.}~\bibnamefont {Brunner}},\ and\ \bibinfo {author} {\bibfnamefont {H.}~\bibnamefont {Scharfetter}},\ }\href {https://doi.org/10.1088/0967-3334/26/2/023} {\bibfield  {journal} {\bibinfo  {journal} {Physiological Measurement}\ }\textbf {\bibinfo {volume} {26}},\ \bibinfo {pages} {S241} (\bibinfo {year} {2005})}\BibitemShut {NoStop}%
\bibitem [{\citenamefont {Hämäläinen}\ \emph {et~al.}(1993)\citenamefont {Hämäläinen}, \citenamefont {Hari}, \citenamefont {Ilmoniemi}, \citenamefont {Knuutila},\ and\ \citenamefont {Lounasmaa}}]{hamalainen_magnetoencephalography---theory_1993}%
  \BibitemOpen
  \bibfield  {author} {\bibinfo {author} {\bibfnamefont {M.}~\bibnamefont {Hämäläinen}}, \bibinfo {author} {\bibfnamefont {R.}~\bibnamefont {Hari}}, \bibinfo {author} {\bibfnamefont {R.~J.}\ \bibnamefont {Ilmoniemi}}, \bibinfo {author} {\bibfnamefont {J.}~\bibnamefont {Knuutila}},\ and\ \bibinfo {author} {\bibfnamefont {O.~V.}\ \bibnamefont {Lounasmaa}},\ }\href {https://doi.org/10.1103/RevModPhys.65.413} {\bibfield  {journal} {\bibinfo  {journal} {Reviews of Modern Physics}\ }\textbf {\bibinfo {volume} {65}},\ \bibinfo {pages} {413} (\bibinfo {year} {1993})},\ \bibinfo {note} {publisher: American Physical Society}\BibitemShut {NoStop}%
\bibitem [{\citenamefont {Alvarez}(1990)}]{alvarez_biomagnetic_1990}%
  \BibitemOpen
  \bibfield  {author} {\bibinfo {author} {\bibfnamefont {R.}~\bibnamefont {Alvarez}},\ }\href {https://doi.org/10.1109/42.57767} {\bibfield  {journal} {\bibinfo  {journal} {IEEE Transactions on Medical Imaging}\ }\textbf {\bibinfo {volume} {9}},\ \bibinfo {pages} {299} (\bibinfo {year} {1990})},\ \bibinfo {note} {conference Name: IEEE Transactions on Medical Imaging}\BibitemShut {NoStop}%
\bibitem [{\citenamefont {Knappe}\ \emph {et~al.}(2016)\citenamefont {Knappe}, \citenamefont {Alem}, \citenamefont {Sheng},\ and\ \citenamefont {Kitching}}]{knappe_microfabricated_2016}%
  \BibitemOpen
  \bibfield  {author} {\bibinfo {author} {\bibfnamefont {S.}~\bibnamefont {Knappe}}, \bibinfo {author} {\bibfnamefont {O.}~\bibnamefont {Alem}}, \bibinfo {author} {\bibfnamefont {D.}~\bibnamefont {Sheng}},\ and\ \bibinfo {author} {\bibfnamefont {J.}~\bibnamefont {Kitching}},\ }\href {https://doi.org/10.1088/1742-6596/723/1/012055} {\bibfield  {journal} {\bibinfo  {journal} {Journal of Physics: Conference Series}\ }\textbf {\bibinfo {volume} {723}},\ \bibinfo {pages} {012055} (\bibinfo {year} {2016})},\ \bibinfo {note} {publisher: IOP Publishing}\BibitemShut {NoStop}%
\bibitem [{\citenamefont {Limes}\ \emph {et~al.}(2020)\citenamefont {Limes}, \citenamefont {Foley}, \citenamefont {Kornack}, \citenamefont {Caliga}, \citenamefont {McBride}, \citenamefont {Braun}, \citenamefont {Lee}, \citenamefont {Lucivero},\ and\ \citenamefont {Romalis}}]{limes_portable_2020}%
  \BibitemOpen
  \bibfield  {author} {\bibinfo {author} {\bibfnamefont {M.}~\bibnamefont {Limes}}, \bibinfo {author} {\bibfnamefont {E.}~\bibnamefont {Foley}}, \bibinfo {author} {\bibfnamefont {T.}~\bibnamefont {Kornack}}, \bibinfo {author} {\bibfnamefont {S.}~\bibnamefont {Caliga}}, \bibinfo {author} {\bibfnamefont {S.}~\bibnamefont {McBride}}, \bibinfo {author} {\bibfnamefont {A.}~\bibnamefont {Braun}}, \bibinfo {author} {\bibfnamefont {W.}~\bibnamefont {Lee}}, \bibinfo {author} {\bibfnamefont {V.}~\bibnamefont {Lucivero}},\ and\ \bibinfo {author} {\bibfnamefont {M.}~\bibnamefont {Romalis}},\ }\href {https://doi.org/10.1103/PhysRevApplied.14.011002} {\bibfield  {journal} {\bibinfo  {journal} {Physical Review Applied}\ }\textbf {\bibinfo {volume} {14}},\ \bibinfo {pages} {011002} (\bibinfo {year} {2020})},\ \bibinfo {note} {publisher: American Physical Society}\BibitemShut {NoStop}%
\bibitem [{\citenamefont {Roth}\ \emph {et~al.}(1989)\citenamefont {Roth}, \citenamefont {Sepulveda},\ and\ \citenamefont {Wikswo}}]{roth_using_1989}%
  \BibitemOpen
  \bibfield  {author} {\bibinfo {author} {\bibfnamefont {B.~J.}\ \bibnamefont {Roth}}, \bibinfo {author} {\bibfnamefont {N.~G.}\ \bibnamefont {Sepulveda}},\ and\ \bibinfo {author} {\bibfnamefont {J.~P.}\ \bibnamefont {Wikswo}},\ }\href {https://doi.org/10.1063/1.342549} {\bibfield  {journal} {\bibinfo  {journal} {Journal of Applied Physics}\ }\textbf {\bibinfo {volume} {65}},\ \bibinfo {pages} {361} (\bibinfo {year} {1989})}\BibitemShut {NoStop}%
\bibitem [{\citenamefont {Levine}\ \emph {et~al.}(2019)\citenamefont {Levine}, \citenamefont {Turner}, \citenamefont {Kehayias}, \citenamefont {Hart}, \citenamefont {Langellier}, \citenamefont {Trubko}, \citenamefont {Glenn}, \citenamefont {Fu},\ and\ \citenamefont {Walsworth}}]{levine_principles_2019}%
  \BibitemOpen
  \bibfield  {author} {\bibinfo {author} {\bibfnamefont {E.~V.}\ \bibnamefont {Levine}}, \bibinfo {author} {\bibfnamefont {M.~J.}\ \bibnamefont {Turner}}, \bibinfo {author} {\bibfnamefont {P.}~\bibnamefont {Kehayias}}, \bibinfo {author} {\bibfnamefont {C.~A.}\ \bibnamefont {Hart}}, \bibinfo {author} {\bibfnamefont {N.}~\bibnamefont {Langellier}}, \bibinfo {author} {\bibfnamefont {R.}~\bibnamefont {Trubko}}, \bibinfo {author} {\bibfnamefont {D.~R.}\ \bibnamefont {Glenn}}, \bibinfo {author} {\bibfnamefont {R.~R.}\ \bibnamefont {Fu}},\ and\ \bibinfo {author} {\bibfnamefont {R.~L.}\ \bibnamefont {Walsworth}},\ }\href {https://doi.org/doi:10.1515/nanoph-2019-0209} {\bibfield  {journal} {\bibinfo  {journal} {Nanophotonics}\ }\textbf {\bibinfo {volume} {8}},\ \bibinfo {pages} {1945} (\bibinfo {year} {2019})}\BibitemShut {NoStop}%
\bibitem [{\citenamefont {Jin}\ \emph {et~al.}(2017)\citenamefont {Jin}, \citenamefont {McCann}, \citenamefont {Froustey},\ and\ \citenamefont {Unser}}]{jin_deep_2017}%
  \BibitemOpen
  \bibfield  {author} {\bibinfo {author} {\bibfnamefont {K.~H.}\ \bibnamefont {Jin}}, \bibinfo {author} {\bibfnamefont {M.~T.}\ \bibnamefont {McCann}}, \bibinfo {author} {\bibfnamefont {E.}~\bibnamefont {Froustey}},\ and\ \bibinfo {author} {\bibfnamefont {M.}~\bibnamefont {Unser}},\ }\href {https://doi.org/10.1109/TIP.2017.2713099} {\bibfield  {journal} {\bibinfo  {journal} {IEEE Transactions on Image Processing}\ }\textbf {\bibinfo {volume} {26}},\ \bibinfo {pages} {4509} (\bibinfo {year} {2017})},\ \bibinfo {note} {arXiv: 1611.03679}\BibitemShut {NoStop}%
\bibitem [{\citenamefont {Wang}\ \emph {et~al.}(2020)\citenamefont {Wang}, \citenamefont {Ye},\ and\ \citenamefont {De~Man}}]{wang2020deep}%
  \BibitemOpen
  \bibfield  {author} {\bibinfo {author} {\bibfnamefont {G.}~\bibnamefont {Wang}}, \bibinfo {author} {\bibfnamefont {J.~C.}\ \bibnamefont {Ye}},\ and\ \bibinfo {author} {\bibfnamefont {B.}~\bibnamefont {De~Man}},\ }\href {http://dx.doi.org/10.1038/s42256-020-00273-z} {\bibfield  {journal} {\bibinfo  {journal} {Nature Machine Intelligence}\ }\textbf {\bibinfo {volume} {2}},\ \bibinfo {pages} {737} (\bibinfo {year} {2020})}\BibitemShut {NoStop}%
\bibitem [{\citenamefont {Schlemper}\ \emph {et~al.}(2019)\citenamefont {Schlemper}, \citenamefont {Oksuz}, \citenamefont {Clough}, \citenamefont {Duan}, \citenamefont {King}, \citenamefont {Schnabel}, \citenamefont {Hajnal},\ and\ \citenamefont {Rueckert}}]{schlemper2019dautomap}%
  \BibitemOpen
  \bibfield  {author} {\bibinfo {author} {\bibfnamefont {J.}~\bibnamefont {Schlemper}}, \bibinfo {author} {\bibfnamefont {I.}~\bibnamefont {Oksuz}}, \bibinfo {author} {\bibfnamefont {J.~R.}\ \bibnamefont {Clough}}, \bibinfo {author} {\bibfnamefont {J.}~\bibnamefont {Duan}}, \bibinfo {author} {\bibfnamefont {A.~P.}\ \bibnamefont {King}}, \bibinfo {author} {\bibfnamefont {J.~A.}\ \bibnamefont {Schnabel}}, \bibinfo {author} {\bibfnamefont {J.~V.}\ \bibnamefont {Hajnal}},\ and\ \bibinfo {author} {\bibfnamefont {D.}~\bibnamefont {Rueckert}},\ }\href@noop {} {\bibfield  {journal} {\bibinfo  {journal} {arXiv preprint arXiv:1909.10995}\ } (\bibinfo {year} {2019})}\BibitemShut {NoStop}%
\bibitem [{\citenamefont {Zhu}\ \emph {et~al.}(2018)\citenamefont {Zhu}, \citenamefont {Liu}, \citenamefont {Cauley}, \citenamefont {Rosen},\ and\ \citenamefont {Rosen}}]{zhu_image_2018}%
  \BibitemOpen
  \bibfield  {author} {\bibinfo {author} {\bibfnamefont {B.}~\bibnamefont {Zhu}}, \bibinfo {author} {\bibfnamefont {J.~Z.}\ \bibnamefont {Liu}}, \bibinfo {author} {\bibfnamefont {S.~F.}\ \bibnamefont {Cauley}}, \bibinfo {author} {\bibfnamefont {B.~R.}\ \bibnamefont {Rosen}},\ and\ \bibinfo {author} {\bibfnamefont {M.~S.}\ \bibnamefont {Rosen}},\ }\href {https://doi.org/10.1038/nature25988} {\bibfield  {journal} {\bibinfo  {journal} {Nature}\ }\textbf {\bibinfo {volume} {555}},\ \bibinfo {pages} {487} (\bibinfo {year} {2018})},\ \bibinfo {note} {number: 7697 Publisher: Nature Publishing Group}\BibitemShut {NoStop}%
\bibitem [{\citenamefont {Coccorese}\ \emph {et~al.}(1994)\citenamefont {Coccorese}, \citenamefont {Martone},\ and\ \citenamefont {Morabito}}]{coccorese_neural_1994}%
  \BibitemOpen
  \bibfield  {author} {\bibinfo {author} {\bibfnamefont {E.}~\bibnamefont {Coccorese}}, \bibinfo {author} {\bibfnamefont {R.}~\bibnamefont {Martone}},\ and\ \bibinfo {author} {\bibfnamefont {F.}~\bibnamefont {Morabito}},\ }\href {https://doi.org/10.1109/20.312527} {\bibfield  {journal} {\bibinfo  {journal} {IEEE Transactions on Magnetics}\ }\textbf {\bibinfo {volume} {30}},\ \bibinfo {pages} {2829} (\bibinfo {year} {1994})},\ \bibinfo {note} {conference Name: IEEE Transactions on Magnetics}\BibitemShut {NoStop}%
\bibitem [{\citenamefont {Kishimoto}\ \emph {et~al.}(1996)\citenamefont {Kishimoto}, \citenamefont {Sakasai},\ and\ \citenamefont {Ara}}]{kishimoto_solution_1996}%
  \BibitemOpen
  \bibfield  {author} {\bibinfo {author} {\bibfnamefont {M.}~\bibnamefont {Kishimoto}}, \bibinfo {author} {\bibfnamefont {K.}~\bibnamefont {Sakasai}},\ and\ \bibinfo {author} {\bibfnamefont {K.}~\bibnamefont {Ara}},\ }\href {https://doi.org/10.1063/1.360946} {\bibfield  {journal} {\bibinfo  {journal} {Journal of Applied Physics}\ }\textbf {\bibinfo {volume} {79}},\ \bibinfo {pages} {1} (\bibinfo {year} {1996})}\BibitemShut {NoStop}%
\bibitem [{\citenamefont {McCann}\ \emph {et~al.}(2017)\citenamefont {McCann}, \citenamefont {Jin},\ and\ \citenamefont {Unser}}]{mccann_convolutional_2017}%
  \BibitemOpen
  \bibfield  {author} {\bibinfo {author} {\bibfnamefont {M.~T.}\ \bibnamefont {McCann}}, \bibinfo {author} {\bibfnamefont {K.~H.}\ \bibnamefont {Jin}},\ and\ \bibinfo {author} {\bibfnamefont {M.}~\bibnamefont {Unser}},\ }\href {https://doi.org/10.1109/MSP.2017.2739299} {\bibfield  {journal} {\bibinfo  {journal} {IEEE Signal Processing Magazine}\ }\textbf {\bibinfo {volume} {34}},\ \bibinfo {pages} {85} (\bibinfo {year} {2017})},\ \bibinfo {note} {conference Name: IEEE Signal Processing Magazine}\BibitemShut {NoStop}%
\bibitem [{\citenamefont {Dubois}\ \emph {et~al.}(2022)\citenamefont {Dubois}, \citenamefont {Broadway}, \citenamefont {Stark}, \citenamefont {Tschudin}, \citenamefont {Healey}, \citenamefont {Huber}, \citenamefont {Tetienne}, \citenamefont {Greplova},\ and\ \citenamefont {Maletinsky}}]{dubois_untrained_2022}%
  \BibitemOpen
  \bibfield  {author} {\bibinfo {author} {\bibfnamefont {A.}~\bibnamefont {Dubois}}, \bibinfo {author} {\bibfnamefont {D.}~\bibnamefont {Broadway}}, \bibinfo {author} {\bibfnamefont {A.}~\bibnamefont {Stark}}, \bibinfo {author} {\bibfnamefont {M.}~\bibnamefont {Tschudin}}, \bibinfo {author} {\bibfnamefont {A.}~\bibnamefont {Healey}}, \bibinfo {author} {\bibfnamefont {S.}~\bibnamefont {Huber}}, \bibinfo {author} {\bibfnamefont {J.-P.}\ \bibnamefont {Tetienne}}, \bibinfo {author} {\bibfnamefont {E.}~\bibnamefont {Greplova}},\ and\ \bibinfo {author} {\bibfnamefont {P.}~\bibnamefont {Maletinsky}},\ }\href@noop {} {\bibinfo {title} {Untrained physically informed neural network for image reconstruction of magnetic field sources}} (\bibinfo {year} {2022})\BibitemShut {NoStop}%
\bibitem [{\citenamefont {Tschudin}\ \emph {et~al.}(2024)\citenamefont {Tschudin}, \citenamefont {Broadway}, \citenamefont {Siegwolf}, \citenamefont {Schrader}, \citenamefont {Telford}, \citenamefont {Gross}, \citenamefont {Cox}, \citenamefont {Dubois}, \citenamefont {Chica}, \citenamefont {Rama-Eiroa}, \citenamefont {J.~G.~Santos}, \citenamefont {Poggio}, \citenamefont {Ziebel}, \citenamefont {Dean}, \citenamefont {Roy},\ and\ \citenamefont {Maletinsky}}]{tschudin_imaging_2024}%
  \BibitemOpen
  \bibfield  {author} {\bibinfo {author} {\bibfnamefont {M.~A.}\ \bibnamefont {Tschudin}}, \bibinfo {author} {\bibfnamefont {D.~A.}\ \bibnamefont {Broadway}}, \bibinfo {author} {\bibfnamefont {P.}~\bibnamefont {Siegwolf}}, \bibinfo {author} {\bibfnamefont {C.}~\bibnamefont {Schrader}}, \bibinfo {author} {\bibfnamefont {E.~J.}\ \bibnamefont {Telford}}, \bibinfo {author} {\bibfnamefont {B.}~\bibnamefont {Gross}}, \bibinfo {author} {\bibfnamefont {J.}~\bibnamefont {Cox}}, \bibinfo {author} {\bibfnamefont {A.~E.~E.}\ \bibnamefont {Dubois}}, \bibinfo {author} {\bibfnamefont {D.~G.}\ \bibnamefont {Chica}}, \bibinfo {author} {\bibfnamefont {R.}~\bibnamefont {Rama-Eiroa}}, \bibinfo {author} {\bibfnamefont {E.}~\bibnamefont {J.~G.~Santos}}, \bibinfo {author} {\bibfnamefont {M.}~\bibnamefont {Poggio}}, \bibinfo {author} {\bibfnamefont {M.~E.}\ \bibnamefont {Ziebel}}, \bibinfo {author} {\bibfnamefont {C.~R.}\ \bibnamefont {Dean}}, \bibinfo {author} {\bibfnamefont {X.}~\bibnamefont {Roy}},\ and\ \bibinfo {author}
  {\bibfnamefont {P.}~\bibnamefont {Maletinsky}},\ }\href {https://doi.org/10.1038/s41467-024-49717-9} {\bibfield  {journal} {\bibinfo  {journal} {Nature Communications}\ }\textbf {\bibinfo {volume} {15}},\ \bibinfo {pages} {6005} (\bibinfo {year} {2024})},\ \bibinfo {note} {publisher: Nature Publishing Group}\BibitemShut {NoStop}%
\bibitem [{\citenamefont {Ronneberger}\ \emph {et~al.}(2015)\citenamefont {Ronneberger}, \citenamefont {Fischer},\ and\ \citenamefont {Brox}}]{ronneberger_u-net_2015}%
  \BibitemOpen
  \bibfield  {author} {\bibinfo {author} {\bibfnamefont {O.}~\bibnamefont {Ronneberger}}, \bibinfo {author} {\bibfnamefont {P.}~\bibnamefont {Fischer}},\ and\ \bibinfo {author} {\bibfnamefont {T.}~\bibnamefont {Brox}},\ }\href {http://arxiv.org/abs/1505.04597} {\bibfield  {journal} {\bibinfo  {journal} {arXiv:1505.04597 [cs]}\ } (\bibinfo {year} {2015})},\ \bibinfo {note} {arXiv: 1505.04597}\BibitemShut {NoStop}%
\bibitem [{\citenamefont {Mehta}\ \emph {et~al.}(2022)\citenamefont {Mehta}, \citenamefont {Padalia}, \citenamefont {Vora},\ and\ \citenamefont {Mehendale}}]{mehta2022mri}%
  \BibitemOpen
  \bibfield  {author} {\bibinfo {author} {\bibfnamefont {D.}~\bibnamefont {Mehta}}, \bibinfo {author} {\bibfnamefont {D.}~\bibnamefont {Padalia}}, \bibinfo {author} {\bibfnamefont {K.}~\bibnamefont {Vora}},\ and\ \bibinfo {author} {\bibfnamefont {N.}~\bibnamefont {Mehendale}},\ }in\ \href@noop {} {\emph {\bibinfo {booktitle} {2022 5th International Conference on Advances in Science and Technology (ICAST)}}}\ (\bibinfo {organization} {IEEE},\ \bibinfo {year} {2022})\ pp.\ \bibinfo {pages} {306--313}\BibitemShut {NoStop}%
\bibitem [{\citenamefont {Chen}\ \emph {et~al.}(2017)\citenamefont {Chen}, \citenamefont {Zhang}, \citenamefont {Zhang}, \citenamefont {Liao}, \citenamefont {Li}, \citenamefont {Zhou},\ and\ \citenamefont {Wang}}]{chen2017low}%
  \BibitemOpen
  \bibfield  {author} {\bibinfo {author} {\bibfnamefont {H.}~\bibnamefont {Chen}}, \bibinfo {author} {\bibfnamefont {Y.}~\bibnamefont {Zhang}}, \bibinfo {author} {\bibfnamefont {W.}~\bibnamefont {Zhang}}, \bibinfo {author} {\bibfnamefont {P.}~\bibnamefont {Liao}}, \bibinfo {author} {\bibfnamefont {K.}~\bibnamefont {Li}}, \bibinfo {author} {\bibfnamefont {J.}~\bibnamefont {Zhou}},\ and\ \bibinfo {author} {\bibfnamefont {G.}~\bibnamefont {Wang}},\ }\href {www.doi-org/10.1364%2FBOE.8.000679} {\bibfield  {journal} {\bibinfo  {journal} {Biomedical Optics Express}\ }\textbf {\bibinfo {volume} {8}},\ \bibinfo {pages} {679} (\bibinfo {year} {2017})}\BibitemShut {NoStop}%
\bibitem [{\citenamefont {Bengio}\ \emph {et~al.}(1994)\citenamefont {Bengio}, \citenamefont {Simard},\ and\ \citenamefont {Frasconi}}]{bengio1994learning}%
  \BibitemOpen
  \bibfield  {author} {\bibinfo {author} {\bibfnamefont {Y.}~\bibnamefont {Bengio}}, \bibinfo {author} {\bibfnamefont {P.}~\bibnamefont {Simard}},\ and\ \bibinfo {author} {\bibfnamefont {P.}~\bibnamefont {Frasconi}},\ }\href@noop {} {\bibfield  {journal} {\bibinfo  {journal} {IEEE transactions on neural networks}\ }\textbf {\bibinfo {volume} {5}},\ \bibinfo {pages} {157} (\bibinfo {year} {1994})}\BibitemShut {NoStop}%
\bibitem [{\citenamefont {Zanforlin}\ \emph {et~al.}(2014)\citenamefont {Zanforlin}, \citenamefont {Munaretto}, \citenamefont {Zanella},\ and\ \citenamefont {Zorzi}}]{zanforlin_ssim-based_2014}%
  \BibitemOpen
  \bibfield  {author} {\bibinfo {author} {\bibfnamefont {M.}~\bibnamefont {Zanforlin}}, \bibinfo {author} {\bibfnamefont {D.}~\bibnamefont {Munaretto}}, \bibinfo {author} {\bibfnamefont {A.}~\bibnamefont {Zanella}},\ and\ \bibinfo {author} {\bibfnamefont {M.}~\bibnamefont {Zorzi}},\ }in\ \href {https://doi.org/10.1109/WIOPT.2014.6850361} {\emph {\bibinfo {booktitle} {2014 12th {International} {Symposium} on {Modeling} and {Optimization} in {Mobile}, {Ad} {Hoc}, and {Wireless} {Networks} ({WiOpt})}}}\ (\bibinfo  {publisher} {IEEE},\ \bibinfo {address} {Hammamet, Tunisia},\ \bibinfo {year} {2014})\ pp.\ \bibinfo {pages} {656--661}\BibitemShut {NoStop}%
\bibitem [{\citenamefont {Flynn}\ \emph {et~al.}(2013)\citenamefont {Flynn}, \citenamefont {Ward}, \citenamefont {Abich},\ and\ \citenamefont {Poole}}]{flynn_image_2013}%
  \BibitemOpen
  \bibfield  {author} {\bibinfo {author} {\bibfnamefont {J.~R.}\ \bibnamefont {Flynn}}, \bibinfo {author} {\bibfnamefont {S.}~\bibnamefont {Ward}}, \bibinfo {author} {\bibfnamefont {J.}~\bibnamefont {Abich}},\ and\ \bibinfo {author} {\bibfnamefont {D.}~\bibnamefont {Poole}},\ }in\ \href {https://doi.org/10.1007/978-3-642-39360-0_3} {\emph {\bibinfo {booktitle} {Engineering {Psychology} and {Cognitive} {Ergonomics}. {Understanding} {Human} {Cognition}}}},\ \bibinfo {series and number} {Lecture {Notes} in {Computer} {Science}},\ \bibinfo {editor} {edited by\ \bibinfo {editor} {\bibfnamefont {D.}~\bibnamefont {Harris}}}\ (\bibinfo  {publisher} {Springer},\ \bibinfo {address} {Berlin, Heidelberg},\ \bibinfo {year} {2013})\ pp.\ \bibinfo {pages} {23--30}\BibitemShut {NoStop}%
\bibitem [{\citenamefont {Hodson}(2022)}]{hodson_root-mean-square_2022}%
  \BibitemOpen
  \bibfield  {author} {\bibinfo {author} {\bibfnamefont {T.~O.}\ \bibnamefont {Hodson}},\ }\href {https://doi.org/10.5194/gmd-15-5481-2022} {\bibfield  {journal} {\bibinfo  {journal} {Geoscientific Model Development}\ }\textbf {\bibinfo {volume} {15}},\ \bibinfo {pages} {5481} (\bibinfo {year} {2022})}\BibitemShut {NoStop}%
\bibitem [{\citenamefont {Winkler}\ and\ \citenamefont {Mohandas}(2008)}]{Winkler_Evolution_2008}%
  \BibitemOpen
  \bibfield  {author} {\bibinfo {author} {\bibfnamefont {S.}~\bibnamefont {Winkler}}\ and\ \bibinfo {author} {\bibfnamefont {P.}~\bibnamefont {Mohandas}},\ }\href {https://doi.org/10.1109/TBC.2008.2000733} {\bibfield  {journal} {\bibinfo  {journal} {IEEE Transactions on Broadcasting}\ }\textbf {\bibinfo {volume} {54}},\ \bibinfo {pages} {660} (\bibinfo {year} {2008})}\BibitemShut {NoStop}%
\bibitem [{\citenamefont {Hauer}\ \emph {et~al.}(2005)\citenamefont {Hauer}, \citenamefont {Kühn},\ and\ \citenamefont {Potthast}}]{hauer_uniqueness_2005}%
  \BibitemOpen
  \bibfield  {author} {\bibinfo {author} {\bibfnamefont {K.-H.}\ \bibnamefont {Hauer}}, \bibinfo {author} {\bibfnamefont {L.}~\bibnamefont {Kühn}},\ and\ \bibinfo {author} {\bibfnamefont {R.}~\bibnamefont {Potthast}},\ }\href {https://doi.org/10.1088/0266-5611/21/3/010} {\bibfield  {journal} {\bibinfo  {journal} {Inverse Problems}\ }\textbf {\bibinfo {volume} {21}},\ \bibinfo {pages} {955} (\bibinfo {year} {2005})}\BibitemShut {NoStop}%
\bibitem [{\citenamefont {"Hinton}\ \emph {et~al.}(2018)\citenamefont {"Hinton}, \citenamefont {Srivastava},\ and\ \citenamefont {Swersky}}]{hinton2018lecture}%
  \BibitemOpen
  \bibfield  {author} {\bibinfo {author} {\bibfnamefont {G.}~\bibnamefont {"Hinton}}, \bibinfo {author} {\bibfnamefont {N.}~\bibnamefont {Srivastava}},\ and\ \bibinfo {author} {\bibfnamefont {K.}~\bibnamefont {Swersky}},\ }\href {"https://www.cs.toronto.edu/~tijmen/csc321/slides/lecture_slides_lec6.pdf"} {\bibinfo {title} {"rmsprop: Divide the gradient by a running average of its recent magnitude (neural networks for machine learning)."}} (\bibinfo {year} {2018})\BibitemShut {NoStop}%
\bibitem [{\citenamefont {Abadi}\ \emph {et~al.}(2016)\citenamefont {Abadi}, \citenamefont {Agarwal}, \citenamefont {Barham}, \citenamefont {Brevdo}, \citenamefont {Chen}, \citenamefont {Citro}, \citenamefont {Corrado}, \citenamefont {Davis}, \citenamefont {Dean}, \citenamefont {Devin} \emph {et~al.}}]{abadi2016tensorflow}%
  \BibitemOpen
  \bibfield  {author} {\bibinfo {author} {\bibfnamefont {M.}~\bibnamefont {Abadi}}, \bibinfo {author} {\bibfnamefont {A.}~\bibnamefont {Agarwal}}, \bibinfo {author} {\bibfnamefont {P.}~\bibnamefont {Barham}}, \bibinfo {author} {\bibfnamefont {E.}~\bibnamefont {Brevdo}}, \bibinfo {author} {\bibfnamefont {Z.}~\bibnamefont {Chen}}, \bibinfo {author} {\bibfnamefont {C.}~\bibnamefont {Citro}}, \bibinfo {author} {\bibfnamefont {G.~S.}\ \bibnamefont {Corrado}}, \bibinfo {author} {\bibfnamefont {A.}~\bibnamefont {Davis}}, \bibinfo {author} {\bibfnamefont {J.}~\bibnamefont {Dean}}, \bibinfo {author} {\bibfnamefont {M.}~\bibnamefont {Devin}}, \emph {et~al.},\ }\href@noop {} {\bibfield  {journal} {\bibinfo  {journal} {arXiv preprint arXiv:1603.04467}\ } (\bibinfo {year} {2016})}\BibitemShut {NoStop}%
\bibitem [{\citenamefont {Broadway}\ \emph {et~al.}(2020)\citenamefont {Broadway}, \citenamefont {Lillie}, \citenamefont {Scholten}, \citenamefont {Rohner}, \citenamefont {Dontschuk}, \citenamefont {Maletinsky}, \citenamefont {Tetienne},\ and\ \citenamefont {Hollenberg}}]{broadway_improved_2020}%
  \BibitemOpen
  \bibfield  {author} {\bibinfo {author} {\bibfnamefont {D.}~\bibnamefont {Broadway}}, \bibinfo {author} {\bibfnamefont {S.}~\bibnamefont {Lillie}}, \bibinfo {author} {\bibfnamefont {S.}~\bibnamefont {Scholten}}, \bibinfo {author} {\bibfnamefont {D.}~\bibnamefont {Rohner}}, \bibinfo {author} {\bibfnamefont {N.}~\bibnamefont {Dontschuk}}, \bibinfo {author} {\bibfnamefont {P.}~\bibnamefont {Maletinsky}}, \bibinfo {author} {\bibfnamefont {J.-P.}\ \bibnamefont {Tetienne}},\ and\ \bibinfo {author} {\bibfnamefont {L.}~\bibnamefont {Hollenberg}},\ }\href {https://link.aps.org/doi/10.1103/PhysRevApplied.14.024076} {\bibfield  {journal} {\bibinfo  {journal} {Physical Review Applied}\ }\textbf {\bibinfo {volume} {14}},\ \bibinfo {pages} {024076} (\bibinfo {year} {2020})}\BibitemShut {NoStop}%
\bibitem [{\citenamefont {Reed}\ \emph {et~al.}(2025)\citenamefont {Reed}, \citenamefont {Bhutto}, \citenamefont {Turner}, \citenamefont {Daly}, \citenamefont {Oliver}, \citenamefont {Tang}, \citenamefont {Olsson}, \citenamefont {Ku}, \citenamefont {Rosen},\ and\ \citenamefont {Walsworth}}]{Reed_Replication_2025}%
  \BibitemOpen
  \bibfield  {author} {\bibinfo {author} {\bibfnamefont {N.}~\bibnamefont {Reed}}, \bibinfo {author} {\bibfnamefont {D.}~\bibnamefont {Bhutto}}, \bibinfo {author} {\bibfnamefont {M.}~\bibnamefont {Turner}}, \bibinfo {author} {\bibfnamefont {D.}~\bibnamefont {Daly}}, \bibinfo {author} {\bibfnamefont {S.}~\bibnamefont {Oliver}}, \bibinfo {author} {\bibfnamefont {J.}~\bibnamefont {Tang}}, \bibinfo {author} {\bibfnamefont {K.}~\bibnamefont {Olsson}}, \bibinfo {author} {\bibfnamefont {M.}~\bibnamefont {Ku}}, \bibinfo {author} {\bibfnamefont {M.}~\bibnamefont {Rosen}},\ and\ \bibinfo {author} {\bibfnamefont {R.}~\bibnamefont {Walsworth}},\ }\href {https://doi.org/10.7910/DVN/SD6PVP} {\bibinfo {title} {{Replication Data for: Machine Learning for Improved Current Density Reconstruction from 2D Vector Magnetic Images}}} (\bibinfo {year} {2025})\BibitemShut {NoStop}%
\end{thebibliography}%

\renewcommand{\thefigure}{S\arabic{figure}}
\setcounter{figure}{0}

\clearpage
\section*{Supplemental Information}
\renewcommand{\thefigure}{S\arabic{figure}}
\setcounter{figure}{0}

\subsection{Experimental Data}
\label{SI_exp}
\begin{figure}[h]
    \centering
    \includegraphics[width=0.5\textwidth]{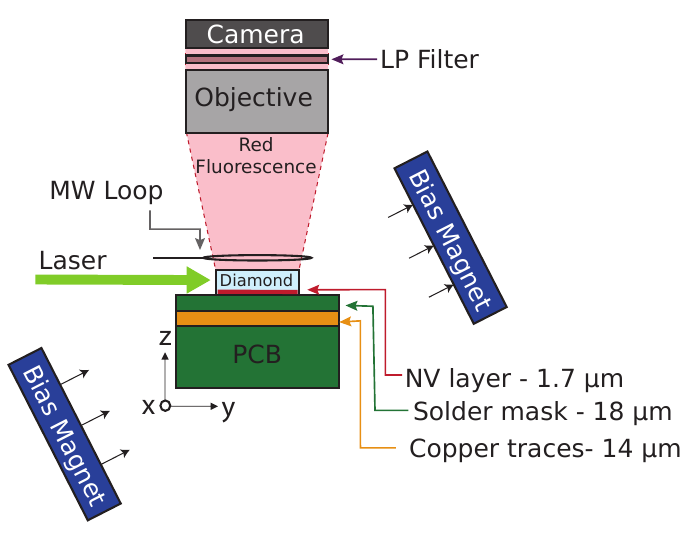}
    \caption{Schematic of experimental set-up. Magnetic field imaging is performed by a quantum diamond microscope (QDM) employing Continuous Wave Optically-Detected Magnetic Resonance (CW-ODMR) measurements of the NV sensing layer at the surface of a diamond chip. The diamond is placed in close proximity to the custom PCB with the NV layer facing downwards.  A wire loop above the diamond delivers a microwave (MW) signal for CW-ODMR measurements, driving the NV electronic spins.  Permanent magnets are positioned to produce a static bias field $\approx 5\,$mT aligned to spectrally separate the four NV orientations. A green laser optically excites the NV ensemble, producing red photoluminescence focused by an optical objective.  A long pass (LP) filter blocks unwanted green light from the CMOS Camera.  Data is generated by running current through the copper trace layer of the custom PCB, producing magnetic fields imaged by the QDM. Precision stages (not shown) adjust the standoff separation of the planar currents in the PCB from the NV sensing layer.}
    \label{NVdiagram}
\end{figure}

We obtain two sets of QDM experimental data using a Continuous Wave Optically-Detected Magnetic Resonance (CW-ODMR) imaging technique: in one ``signal" data set, the PCB is electrically active and producing magnetic fields; whereas for a second ``reference" data set only the ambient magnetic field environment is recorded.  We then fit signal and reference data with custom fitting software to extract vector magnetic field components at each pixel of the images, subtracting reference data from signal data to remove artifacts, e.g., from ferromagnetic components of the PCB and gradients in the bias magnetic field.

To create different levels of noise in the images, we vary the amount of averaging of the measured CW-ODMR data before the fit is performed to determine magnetic field values per pixel. For high SNR images, we average 400 measurements at each MW frequency of the CW-ODMR spectrum; whereas for the low SNR image we perform the fit on a single sweep of MW frequency with no signal averaging.

The QDM employs a pair of 2 inch SmCo magnets to produce a bias field $\approx$ 5$\,$mT, aligned such that the ODMR spectra of the 4 NV orientations are separated, allowing the reconstruction of vector magnetic field measurements. The diamond, fabricated by Element Six\texttrademark, features a 1.7$\,\mu$m NV layer enriched with nitrogen-15 ([N]=17\,ppm, [NV]=2\,ppm). The diamond is isotopically purified to contain $\sim$ 99.995 \% carbon-12. 

For CW-ODMR measurements, a Stanford Research Systems SG384 signal generator produces a sinusoidal MW waveform resonate with the NV spin transitions. The signal is amplified by a Mini Circuits ZHL-16W-43-S+ high power amplifier and delivered through a custom made 7$\,$mm diameter wire loop located above the diamond. The NV layer is optically excited with 1.5\,W of power supplied by a Lighthouse Photonics Sprout-H-10W laser. A 4x optical objective (Olympus UPlanFL N 0.13 NA), a 633\,nm longpass filter (Semrock LP02-633RU-25), and a CMOS CCD camera (Basler acA1920-155um) form the NV photoluminescence imaging system. A Rigol DG1022U arbitrary waveform generator produces a 100$\,$mV DC signal used to drive currents through copper wire traces in the custom PCB, which is mounted on precision mechanical stages below the diamond. The stages allow adjustments of the standoff distance of the planar currents in the PCB from the NV sensing layer. Figure \ref{NVdiagram} shows a schematic of the experimental set-up.

\subsection{Image SNR}

To determine the signal-to-noise ratio (SNR) of experimental magnetic field images, we first find the maximum field strength along a line cut orthogonal to the direction of a wire segment producing a strong signal. Then, we compute the average signal by repeating this process at each pixel along the entire length of the segment. In some images, such as from the Hilbert curve shown in Figure \ref{tiling}d, the strongest wire segments produce fields 2-3 times greater than the weakest signal segments due to cancellation of magnetic fields near closely spaced wires. The average noise is determined by computing the standard deviation of pixels in a low-signal region of the image. 

\subsection{Noise Analysis}
\label{noise_sec}

\begin{figure*}
    \centering
    \includegraphics[width=1.0\textwidth]{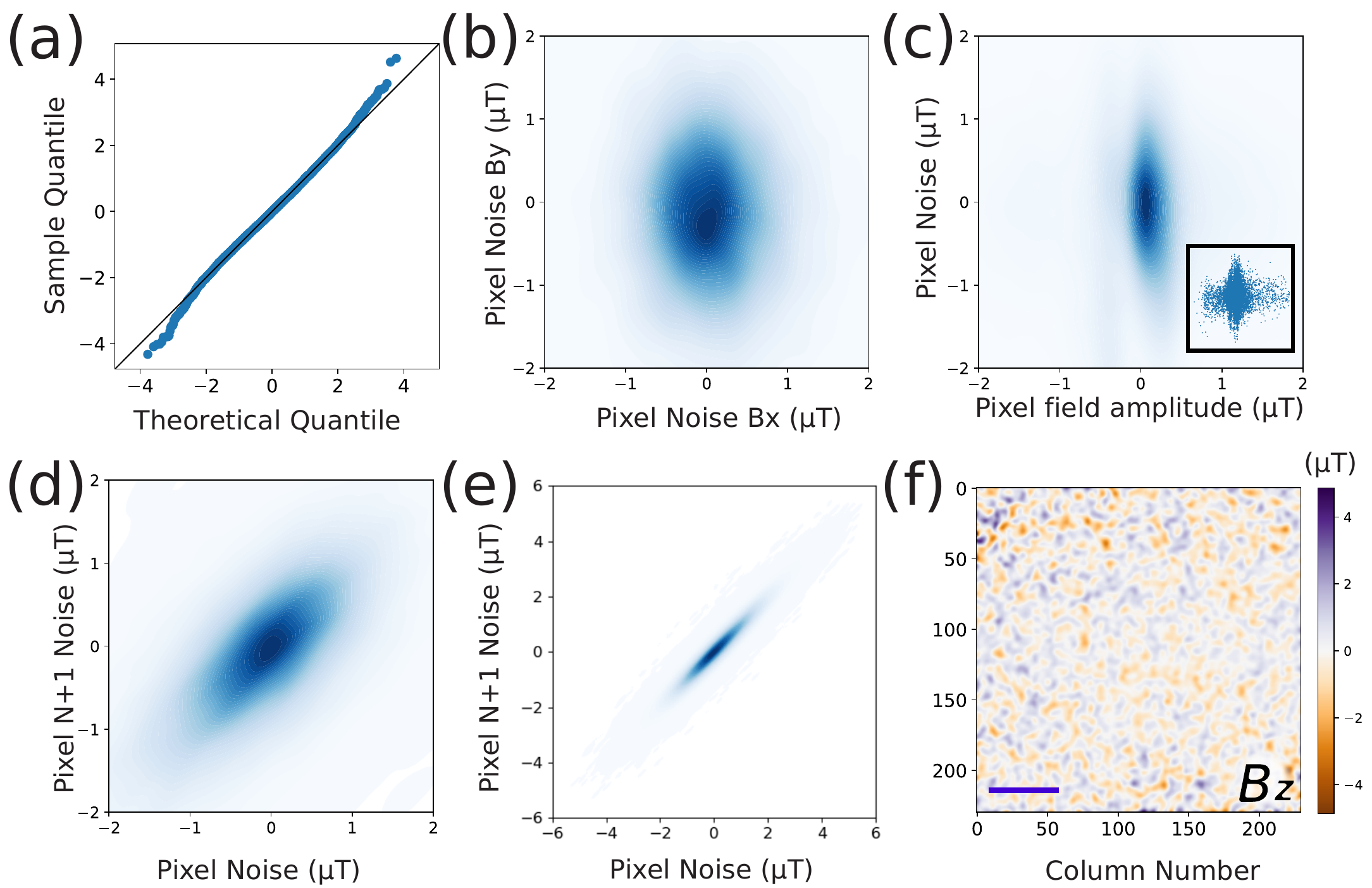}
\caption{Analysis of the noise data set produced from experimental QDM magnetic field images such as those shown in Figures \ref{experimental} and \ref{tiling}.  Seven pairs of low and high SNR images are used to generate the data set, and are compared at 230x230 resolution and at 64x64 resolution after binning and cropping. The noise data set is generated by subtracting high SNR magnetic field images from low SNR images with the same field of view. This noise data set retains vector information, thus consisting of x-noise, y-noise, z-noise images.  (a) Quantile-quantile plot of the noise data set exhibits approximately linear dependence, indicating a normal distribution. (b) Per-pixel x-noise vs per-pixel y-noise of all image pixels. Noise is uncorrelated between vector components. (c) Per-pixel noise vs magnetic field amplitude (of the unprocessed, high-SNR images used to create the noise data set of all image pixels). There is minimal correlation between magnetic field amplitude and noise. Inset shows a scatter plot of this plot for a single image pair to better illustrate the behavior where magnetic field amplitude is large. (d) Noise data set is binned to 64x64 resolution; per pixel amplitude noise for row N, column M pixels plotted against row N+1, column M pixels. There is significant spatial correlation ($R^2=0.60$). (e) Same as (d), binned to 230x230 final resolution. There is larger spatial correlation ($R^2=0.95$). (f) Example $B_z$ component contributing to the noise data set, generated by subtracting a high-SNR iamge from a low-SNR image. Noise artifacts $\approx$ 5-10 pixels can be seen. Scale bar is 400$\,\mu$m for the planar $B_z$ image shown in (f).}
    \label{noise_plots}
\end{figure*}
\noindent

To inform the generation of synthetic data with artificial noise that properly reflects observations in experimental measurements, we analyze the noise characteristics of QDM magnetic field images, such as those shown in Figure \ref{tiling}. For numerous experimental datasets, we subtract high SNR images (constructed from the average of 400 CW-ODMR spectral sweeps with resulting SNR $\approx$ 87) from their low-quality counterparts within the same field of view (a single CW-ODMR sweep with SNR $\approx$ 4.7) to produce noise datasets.  Analysis of these datasets shows that the experimental noise is approximately Gaussian (see Figure \ref{noise_plots}a), with no significant correlation among field components (Figure \ref{noise_plots}b).  This analysis also indicates that the noise is additive, since the noise distribution does not change as a function of field amplitude (Figure \ref{noise_plots}c).

Noise spatial correlations are visually apparent as cluster artifacts in QDM magnetic field images, approximately 5-10 pixels wide (Figure \ref{noise_plots}f); and manifest in Figure \ref{noise_plots}d,e as interdependent variables.  This modest deviation from a purely additive, Gaussian model negatively impacts the performance of the MAGIC-UNet network when applied to experimental data, as the network is trained only on synthetic data utilizing uncorrelated Gaussian noise. For example, MAGIC-UNET performs less well for the 230x230 resolution experimental data set (spatial correlation $R^2=0.95$) than for the 64x64 resolution data (spatial correlation $R^2=0.60$), as seen in Figure \ref{noise_plots}d,e. In future work, these issues may be addressed by (i) reducing spatial correlations in experimental data, whether induced by the hardware or the fitting algorithm; and/or (ii) introducing spatial correlations into the synthetic training data.

\subsection{Synthetic Data Generation}\label{synthetic_data_supplemental}

\begin{figure}[h]
    \centering
    \includegraphics[width=0.5\textwidth]{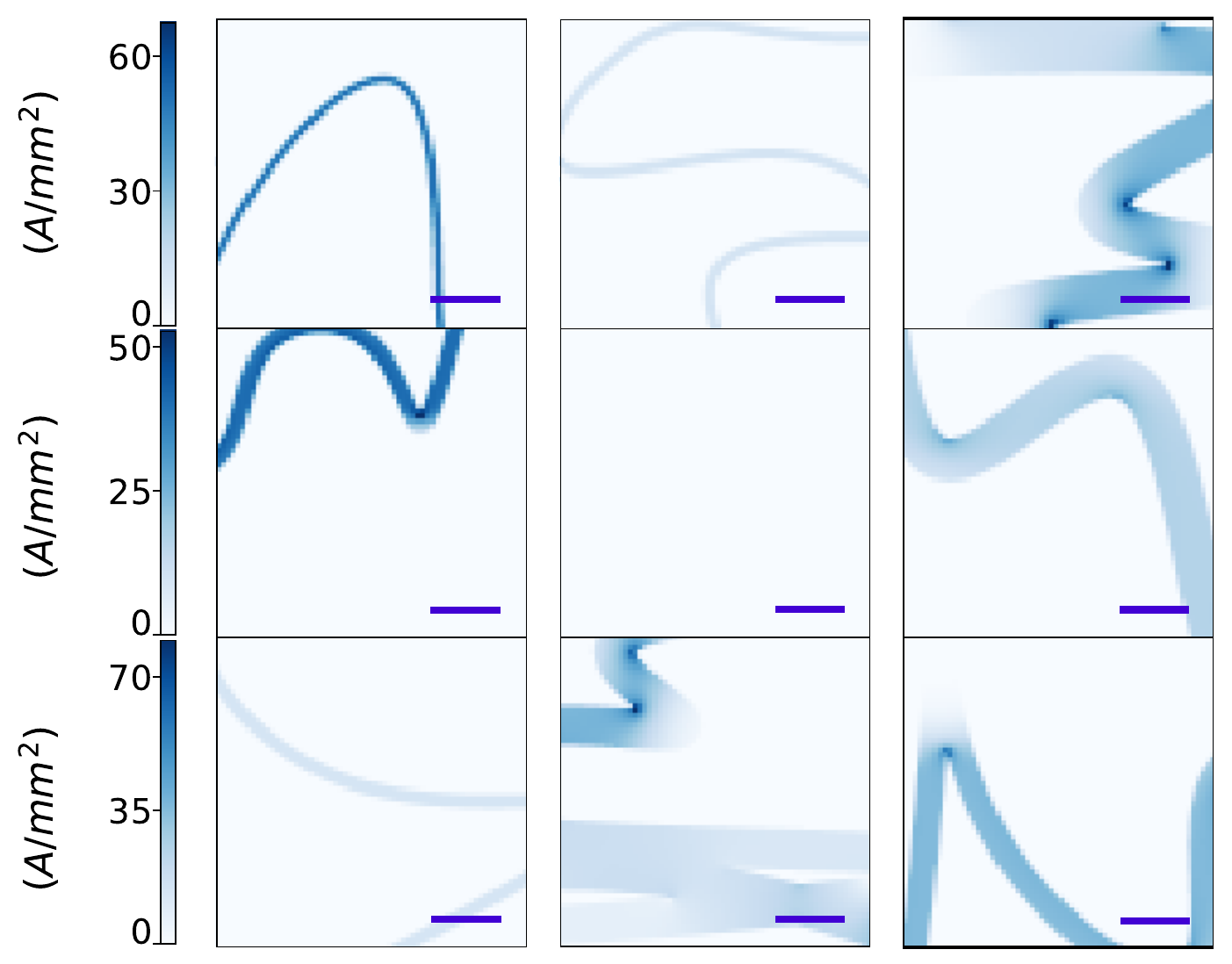}
    \caption{\label{example_data_COMSOL} Examples of current density magnitude from Class I data.  Current flow around bends and overlapping wires is simulated using COMSOL Multiphysics\textsuperscript{\textregistered} to realistically simulate current behavior around corners. Scale bars are 400$\,\mu$m for all planar images shown.}
   
\end{figure}

\begin{figure}[h]
    \centering
    \includegraphics[width=0.5\textwidth]{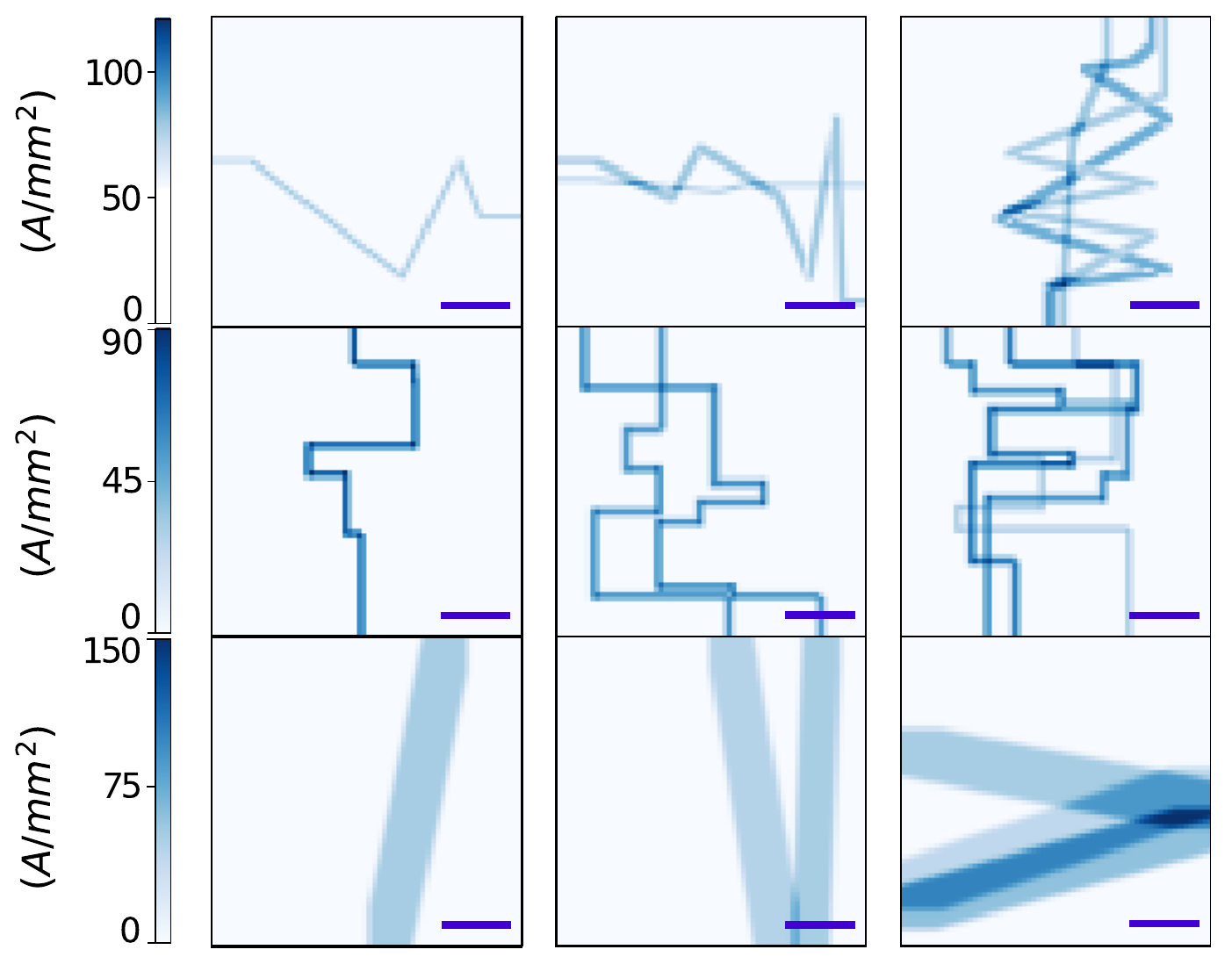}
    \caption{Example current density magnitudes from Class II data. This data includes straight wire segments at different angles, and can include 1, 2, or 3 closely spaced current planes. Scale bars are 400$\,\mu$m for all planar images shown.}
    \label{example_data_MATLAB}
\end{figure}

Synthetic data generation incorporates several key steps to ensure robust training of the MAGIC-UNet. To begin, we generate randomized planar current density images by varying total current, wire widths, numbers of bends, locations and angles of wires, and standoff distances across two classes of wire types.  ``Class I"  features curving and looping wires with up to 20 bends, allowing wires to form shorts if they cross (see Figure \ref{example_data_COMSOL}). This class aids the network in learning about non-uniform current flow. ``Class II" wires consist of only straight segments, changing directions at arbitrary angles and forming random intersections (see Figure \ref{example_data_MATLAB}).  We add current densities together at these intersections to emulate the effect of separated wires crossing in chips featuring insulating layers.  Key parameters used are listed in Table \ref{td_table}. We choose parameters both to approximate common behaviors in our PCB sample and support a robust and unbiased model.

After generating synthetic data, we increase image diversity by employing three separate unitary operations: 90 degree rotation, horizontal reflection, and vertical reflection.  For a given image, each operation occurs with 50\% probability, resulting in 8 possible permutations. This process helps eliminate sources of bias in current density components, directions, and spatial locations.  Additionally, approximately 2\% of the training data features wires exclusively outside of the field of view, in order to teach the network to understand fields generated outside the region of interest.

We then produce a final set of current densities using the Biot-Savart law recast as a Fourier Transform.  This Forward Fourier Method produces a high fidelity solution since the current density images do not have noise added at this stage in the synthetic data generation process. We normalize each input magnetic field image individually to create a more optimal normalization scheme, as the input data has a wide spread of maximum values. Let the scalar used to normalize each image, $i$, be $\alpha_i$.  Corresponding output images are then also scaled by $\alpha_i$ to maintain a proper relationship between magnetic field amplitude and current density, given finite computational precision. The network performs best when a separate, global scaling factor $\beta$ ($\sim10^8$) is applied to all output data before training; and the inverse of this factor $1/\beta$ is applied afterwards to the MAGIC-UNet prediction. Given that $\beta$ affects average values by many orders of magnitude, this operation helps ensure that the mean pixel value of input data is not so small or large as to limit the flexibility of the network's activation function. Between every epoch of the training process, additive Gaussian noise drawn from a random seed is applied to the input data. This noise is found to be a reasonable match to the noise behavior of our experimental data.

To evaluate the performance of the trained MAGIC-UNet network on out-of-distribution data, we create sets of input-output pairs, generating each set with different methods.  In one set, we simulate random, straight-current densities directly in COMSOL\textsuperscript{\textregistered} using an alternative algorithm and with separate parameters to the synthetic Class II data described above. We then obtain predicted magnetic field amplitudes from these current densities using COMSOL\textsuperscript{\textregistered}. These simulated magnetic field images not only serve as effective, out-of-distribution outputs, but also as a method to validate the Forward Fourier Method calculation for creating training data. Separately, we generate custom current densities in COMSOL\textsuperscript{\textregistered} that mimic current flow through devices with branched current and nonuniform wire width to test MAGIC-UNet performance on data far outside of the distribution of training data.

\noindent

\noindent

\renewcommand{\thetable}{S\arabic{table}}
\setcounter{table}{0}

\begin{table*}
\begin{ruledtabular}
\begin{tabular}{c|cc}
Parameter & Class I & Class II \\
\hline
\hline
Class of wire & \specialcell{Curves formed by interpolating 2-21 randomly \\ generated points (with cropping)} & \specialcell{Thin right angle segments, thin arbitrary \\ angle segments, thick straight wires} \\
\hline
Number of independent wires & 1 & 1, 2, or 3 \\
\hline
Field of view & 2$\,$mm & 2$\,$mm \\
\hline
Standoff distance & 50 ± 10$\,\mu$m (500 ± 50$\,\mu$m for large standoff)& 50 ± 10$\,\mu$m (500 ± 50$\,\mu$m for large standoff) \\
\hline
Thickness of current layer & 14 μm & 14 μm \\
\hline
Width of wires (\% of data set)& 16-160$\,\mu$m (50\%)
160-320$\,\mu$m (50\%)& 9-30 μm (67\%)
97-156 μm (33\%)\\
\hline
Approximate Current Range& 15.5 - 311$\,$mA & 1.95 - 170$\,$mA \\
\end{tabular}
\end{ruledtabular}
\caption{\label{td_table}Parameters for synthetic current density data generation, used for MAGIC-UNet training and validation.}
\end{table*}

\end{document}